\documentclass{aa}
\usepackage[utf8]{inputenc}

\usepackage{graphicx}
\usepackage{txfonts}
\usepackage[colorlinks, citecolor=blue, breaklinks=true]{hyperref}
\usepackage{longtable}
\usepackage{siunitx}
\usepackage[normalem]{ulem}
\usepackage{xcolor}
\usepackage{lscape}
\usepackage[font={footnotesize},justification=justified, singlelinecheck=false, labelfont=bf]{caption}

\def\Msun{M$_\odot$}
\def\kms{km~s$^{-1}$}
\def\spin{\mathrm{spin}}
\def\orb{\mathrm{orb}}
\def\sync{\mathrm{sync}}
\def\wind{\mathrm{wind}}
\def\Teff{T_\mathrm{eff}}
\def\veq{$v_\mathrm{eq}$}
\def\tsync{$t_\mathrm{sync}$}

\begin{document}

\title{Ages and ZAMS spin distribution of stars in detached eclipsing binaries}
\author{T. Merle\inst{1,2}, W. Van Rensbergen\inst{3}, L. Siess\inst{1}, J.P. De Greve\inst{3}, K. Jansen\inst{3}, S. Van Eck\inst{1}, G. Van de Steene\inst{2}  }

\institute{
$^1$ BLU-ULB, Institut d'Astronomie et d'Astrophysique, Universit\'e Libre de Bruxelles, CP 226, Bvd du Triomphe,   B-1050 Brussels\\
$^2$ Royal Observatory of Belgium, Avenue Circulaire 3, B-1180 Brussels\\
\email{thibault.merle@ulb.be}\\
$^3$ Astrophysical Institute, Vrije Universiteit Brussel (VUB) , Pleinlaan 2, B-1050 Brussels\\
}
\date{Received month, day, year; accepted month, day, year}
\titlerunning{Ages and ZAMS spins of DEB components}
\authorrunning{T. Merle}

\abstract{
Benchmarking fundamental stellar properties is essential for calibrating evolutionary models and establishing empirical relationships between mass, radius, luminosity and related stellar properties.
We determine the current ages and initial equatorial velocities at the Zero-Age Main Sequence (ZAMS) for a well-characterized sample of 108 detached main-sequence eclipsing binaries.  
Evolutionary tracks from the ZAMS to the present are calculated by accounting for tidal interactions, including meridional circulation, and angular momentum loss via stellar winds, under the assumption of circular orbits. System ages are derived by identifying the evolutionary stage at which the calculated radii of both stellar components best match the observed values. 
While initial velocities for currently synchronized systems cannot be uniquely constrained, as a wide range of initial states naturally converges to synchronisation over time, we are able to determine unique solutions for ZAMS spin velocities in systems that remain asynchronous today. Our models successfully reproduce observed present-day equatorial velocities with a precision of 1\%. Two systems, HD~71636 and V396~Cas, were found to have primaries with initially retrograde spin at ZAMS. 
We also find an increasing dispersion of spin velocities with age. Our results demonstrate that tidal and evolutionary effects in binary systems actively counteract rotational deceleration from stellar winds, effectively preventing the substantial spin-down that typically characterizes the evolution of isolated single stars.
}

\maketitle

\section{Introduction}
Benchmark stars play a pivotal role in modern astrophysics for several key reasons: (i) the cross-calibration of large-scale, ground-based spectroscopic surveys \citep[e.g.,][]{soubiran2024,giribaldi2023,karovicova2022}, (ii) the calibration of stellar evolutionary models \citep[e.g.,][]{martins2013,creevey2024}, and (iii) the establishment of empirical relationships between fundamental stellar properties such as mass, luminosity, radius, and related quantities. \citep{moya2018,eker2024}. Among the most precise and reliable stellar parameters are those derived from eclipsing binaries exhibiting both radial velocity variations and mutual eclipses or transits. Comprehensive compilations of such high-precision parameters are provided in seminal works such as \citet{andersen1991}, \citet{torres2010}, \citet{eker2014}, and \citet{southworth2015}, among others. A detailed review by \citet{serenelli2021} further highlights the exceptional precision achievable in stellar mass determinations, ranging from 2\% for massive stars to better than 0.5\% for low-mass stars. Recent advances, including the use of Gaia astrometry for spectroscopic binary systems (SB2), have significantly improved these efforts \citep[e.g.][]{halbwachs2020,chevalier2023}.

Among stellar properties, age remains one of the most challenging to constrain, as it leaves minimal direct imprints on stellar spectra \citep{soderblom2010}. Isochrone fitting and evolutionary tracks are widely used to estimate stellar ages, although discrepancies can arise between different stellar evolution models. While determining the age of a single star (particularly one on the main sequence) can be ambiguous or even unfeasible, the analysis of stellar binaries offers a distinct advantage. Binaries provide a unique opportunity to derive a consistent age for both components, assuming co-eval formation and similar chemical composition within their natal star-forming region, which is a reasonable hypothesis.

Initial rotation rates are crucial for advancing our understanding of star formation theories  \citep{Rosen2012, Kamann2025}, explaining the contraction from proto-star to ZAMS and angular momentum loss due to stellar winds \citep{douglas2024}, as well as  studying tidal interactions in binary systems \citep{lai1997,vanrensbergen2020}. However, estimating the Zero Age Main Sequence (ZAMS) rotational velocities remains a significant challenge.

In this study, we determine the ages and ZAMS spin distributions of a well-defined subsample of detached, eclipsing binaries from the catalogue of \citet{eker2014} assuming circular orbits. Section~2 outlines the criteria for system selection, while Sections~3 and 4 explore tidal interactions from the ZAMS to the present day and the loss of angular momentum through stellar winds. The derived ages and initial equatorial velocities are presented in Section~5.


\begin{figure*}
    \includegraphics[width=0.33\linewidth]{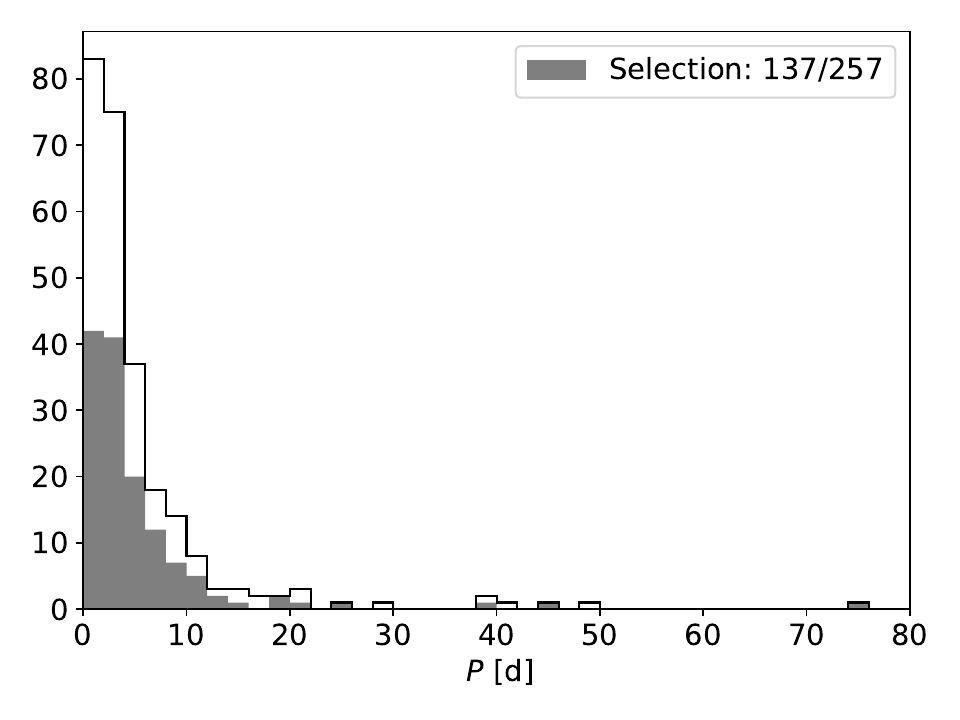}
    \includegraphics[width=0.33\linewidth]{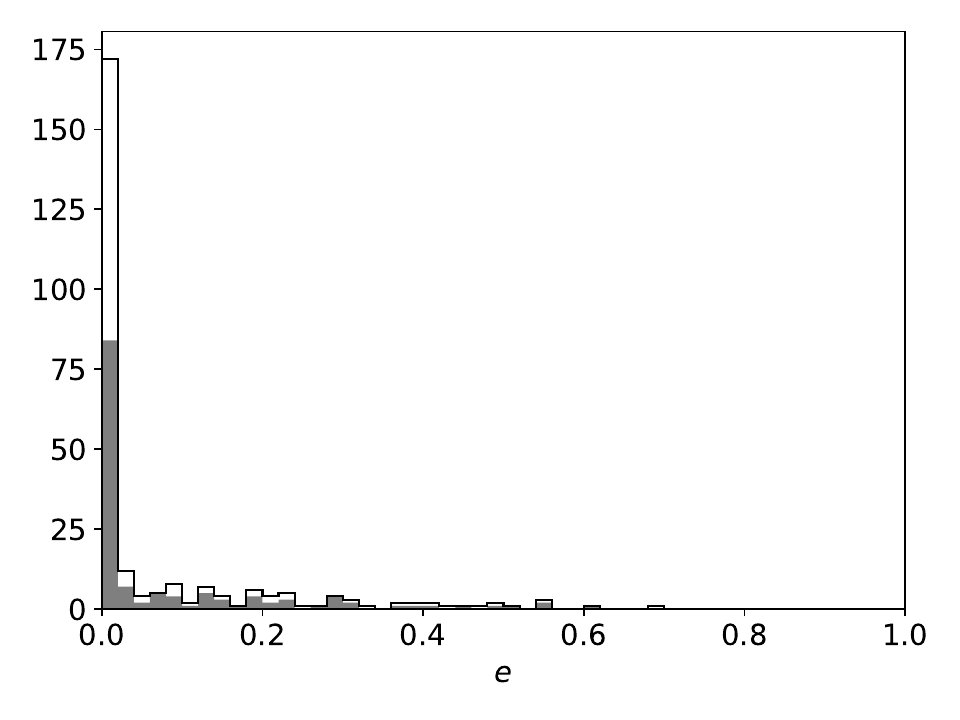}    
    \includegraphics[width=0.33\linewidth]{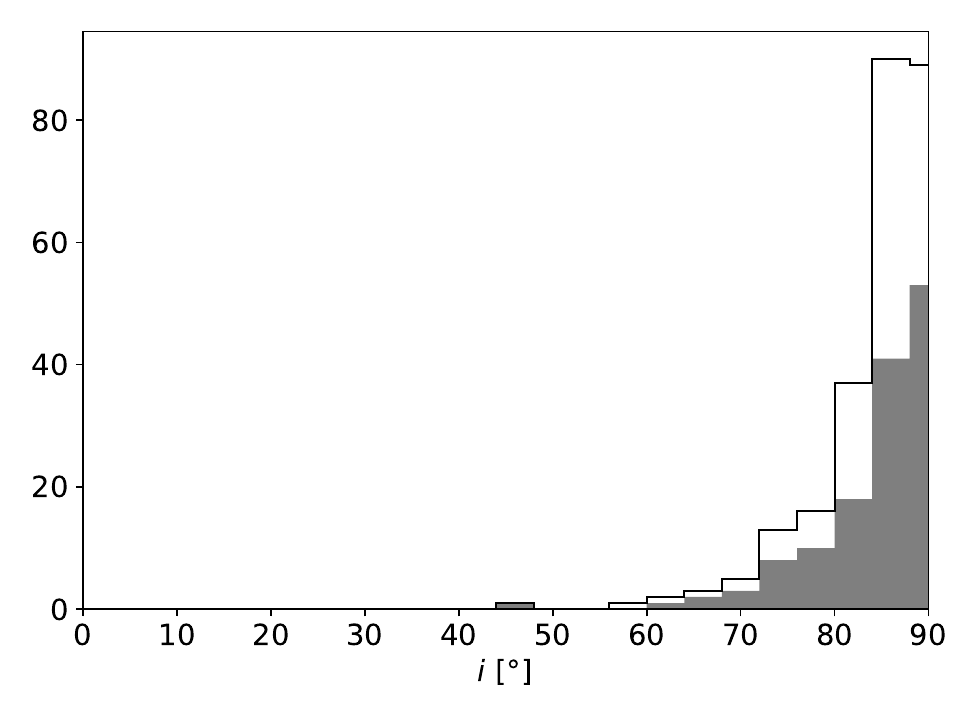}  
    \caption{Periods, eccentricities and inclinations distribution of binary stars from \citet{eker2014} (white histograms) and our selection (grey histograms) having measured values of effective temperatures, masses and projected rotational velocities.}
    \label{fig:histo_pei}
\end{figure*}

\section{Sample Selection}
\label{sect:sample}

The binary systems analyzed in this study are drawn from \emph{The Catalogue of Stellar Parameters from the Detached Double-Lined Eclipsing Binaries in the Milky Way} \citep{eker2014}, with the majority located within 1 kpc of the Sun. This catalogue comprises 257 double-lined, detached spectroscopic binaries, each with at least one component whose mass and radius are determined with a precision better than 3\%. For our analysis, we selected systems with well-defined parameters: effective temperature ($\Teff$), luminosity ($L$), radius ($R$), mass ($M$), and equatorial velocity (\veq) for both components. This selection yielded an initial sample of 137 binaries, with their distributions of orbital periods, eccentricities, and inclinations illustrated in Fig.~\ref{fig:histo_pei}. Approximately 75\% of these systems exhibit eccentricities below 0.1.
As expected, the systems exhibiting the highest eccentricities ($e>0.5$), namely LV~Her, V1143~Cyg, and HW~CMa, are distinctively found in wider orbits (P=18.4, 7.6, and 21.1 days, respectively) where tidal forces remain highly inefficient.

From the initial 137 systems, 24 were excluded for the following reasons: (i) systems containing one or two M-type components for which derived ages are unrealistic, often exceeding the Hubble time, due to the underestimation of their radii by stellar evolution codes, and (ii) the systems TZ~For (comprising two subgiants with the longest orbital period in the sample, 75.7 days) and AL Scl (whose primary has the highest equatorial velocity, \veq$=304$ km s$^{-1}$) due to significant evolution off the main sequence in one of their components.

The final sample consists of 108 systems, with their stellar properties compiled from \citet{eker2014} and listed in Table~\ref{tab:main}. The catalogue provides the orbital inclination angle ($i$) relative to the line of sight and the projected equatorial velocities (\veq$\sin{i}$), enabling us to derive the true equatorial velocities for each component. The system with the smallest inclination (47°) is DH~Cep, while LV~Her exhibits the highest eccentricity (0.613). 

\begin{figure*}
    \centering
    \includegraphics[width=0.49\linewidth]{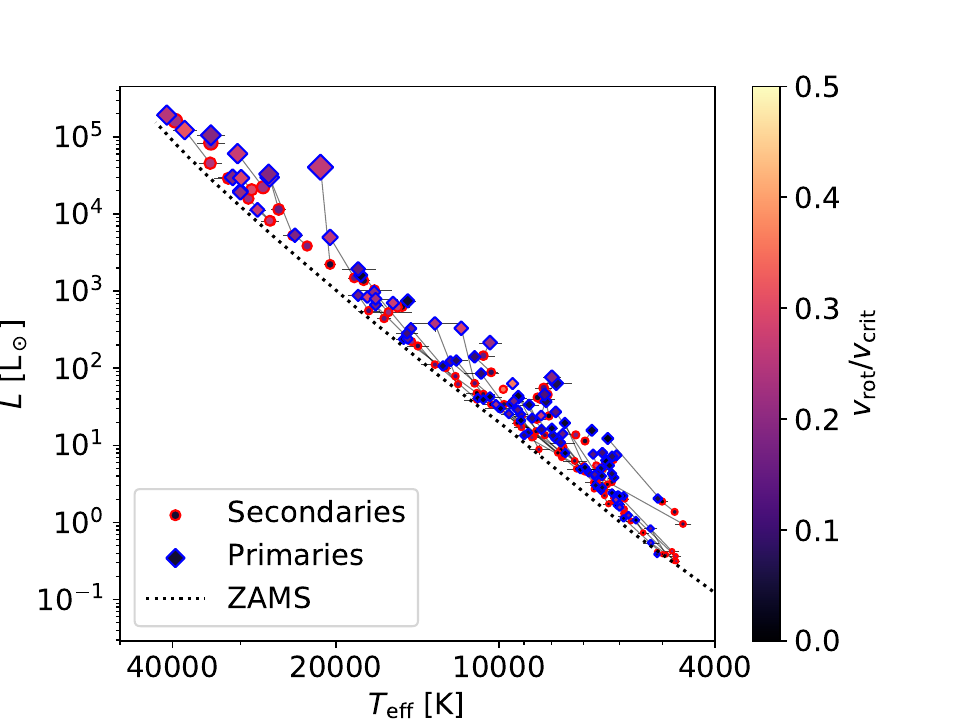}
    \includegraphics[width=0.49\linewidth]{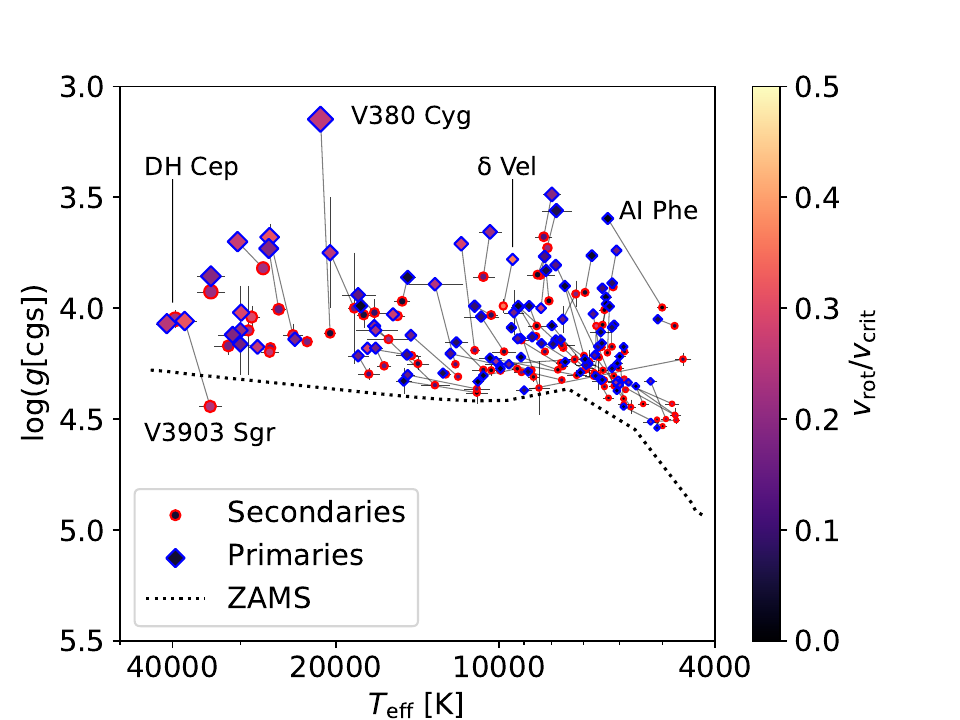}
    \caption{Hertzsprung-Russell (left) and Kiel (right) diagrams of the 108 systems in the final selection of eclipsing binary stars. Components of a given system are linked by gray lines. The size of the symbols scales with the radius of the components. Colour codes the ratio of rotational to critical velocities. 
    }
    \label{fig:hrd_kiel}
\end{figure*}

The Hertzsprung-Russell diagram (HRD) and Kiel diagram for the final sample of 108 systems are presented in Fig.~\ref{fig:hrd_kiel}. Symbols are color-coded by the ratio of the equatorial over the critical velocity. The critical velocity for each component is calculated using the relation $v_{\mathrm{crit}, i}= \sqrt{GM_i/R_i} \simeq 436.68\sqrt{M_i/R_i}$ \kms, where $M_i$ and $R_i$ are the mass and radius of component $i$ in solar units, and $v_{\mathrm{crit}, i}$ is in \kms. 
We assumed a solar metallicity at ZAMS following the distribution described by \cite{grevesse2007}: $X=0.70$, $Y=0.28$, and $Z=0.02$.
We also noticed three systems with effective temperature differences between components exceeding 4000 K: IQ~Per, CM~Lac, and V615~Per, which would probably not be observed as a spectroscopic binaries with composite spectra, since the flux of the primary largely dominates over the secondary. 

We identified an inconsistency in the catalogue of \citet{eker2014} for V885 Cyg. In this system, the primary star is less luminous than the secondary because the original article by \citet{lacy2004} designated the hotter star as the primary, despite it being smaller and consequently less luminous than its companion. \citet{lacy2004} noted that the rotational velocities are synchronized, implying that the larger star should have the higher velocity. However, in the catalogue of \citet{eker2014}, the velocities are not correctly assigned to the respective components. 

Our objective is to determine the equatorial velocities (\veq) at ZAMS for the 108 selected binaries. All systems in this sample are observed prior to potential Roche Lobe Overflow (RLOF), from inspection of the light curves \citep{eker2014}. The primary star (subscript = 1) is defined as the more massive component, while the secondary (subscript = 2) is its lower-mass companion. Since mass transfer has not yet begun, the spin evolution of both stars is governed by the change in radius due to stellar evolution, as well as tidal interactions and angular momentum loss by stellar winds. 

\section{Tidal interaction from ZAMS to now}
Tidal forces in binary stars are responsible for both circularising orbits and synchronising stars. We neglect in this study the effects of tides on the eccentricity and only focus on the tidal dissipation within stars that leads to synchronization of the spins with the orbit.

\subsection{Theory of Darwin}
Tidal evolution governs the synchronisation of a binary star's spin angular velocity, $\omega_\spin$, with its orbital angular velocity, $\omega_\orb$, over a characteristic timescale, \tsync. 
The fundamental theory of tidal friction was initially applied to the Earth-Moon system by \citet{darwin1879}, yielding a synchronisation time, which scales as:
\begin{equation}
t_\sync^\mathrm{D} \propto \frac{1}{q^2} \left( \frac{a}{R} \right)^6,
\end{equation}
where $a$ is the separation between the two components, and $q$ is the mass ratio, with the component of interest in the denominator, and the superscript "D" refers to Darwin.
The theory of \citet{darwin1879} was later extended to close binary stars by \citet{zahn1966}\footnote{The weak friction theory for tidal interactions in binary stars has been further developed by \citet{alexander1973}, \citet{kopal1978}, and \citet{hut1981}. More recent developments are from \citet{zahn1989}.}. Sections~\ref{sect:darwing_ce} and \ref{sect:darwing_re} present refinements to Darwin's theory for tidal interactions in binary systems. 

For a star with moment of inertia $I$ rotating asynchronously with its orbit, the change in its spin angular momentum, $J_\spin$, over a time interval $\Delta t$ is given by:
\begin{equation}
\Delta J_\spin = I \left( \omega_\orb - \omega_\spin \right) \left( 1 - e^{-\Delta t / t_\sync} \right).
\end{equation}
The sign of $\Delta J_\spin$ depends on the relative values of $\omega_\orb$ and $\omega_\spin$. If $\omega_\spin > \omega_\orb$, the star's rotation slows down as angular momentum is transferred to the orbit. Conversely, if $\omega_\spin < \omega_\orb$, the star's rotation accelerates to synchronize with the orbit.

\begin{table*}
\centering
\scriptsize
\caption{108 selected detached EB systems (out of 257) from \citet{eker2014} for determining current ages and initial spins.}
\label{tab:main}
\begin{tabular}{llrlrllllrrrr}
\hline
System & Type & $P$  & $e$ & $i$  &$M_1$ &$M_2$&$R_1$ &$R_2$ &$T_{\mathrm{eff},1}$ &$T_{\mathrm{eff},2}$& \veq$_{,1}$ & \veq$_{,2}$ \\ 
 & & [d] & & [$^\circ$] & [\Msun] &  [\Msun] & [$R_\odot$] & [$R_\odot$] & [K] & [K] & [\kms] & [\kms]\\
\hline
UV Psc&G5V+K3V&0.9&0.0&88.9&0.975&0.76&1.11&0.83&5780&4750&68.3&58.3 \\
GZ Leo&K1V+K1V&1.5&0.007&75.0&0.79&0.78&0.79&0.82&5120&5120&27.2&27.9 \\
CG Cyg&G9V+K3V&0.6&0.0&81.4&0.94&0.81&0.89&0.84&5260&4720&80.9&66.8 \\
V565 Lyr&G+G&18.8&0.02&89.4&0.996&0.929&1.101&0.971&5600&5430&4.0&3.5 \\
V1430 Aql&G2V+K4V&0.9&0.0&87.6&0.96&0.86&1.1&0.85&5262&4930&63.6&49.5 \\
NGC188 KR V12&G+G&6.5&0.0&88.6&1.103&1.081&1.424&1.373&5900&5875&15.0&17.0 \\
AI Phe&F7V+K0IV&24.6&0.186&88.4&1.236&1.195&2.93&1.816&6310&5010&6.0&4.0 \\
V636 Cen&G1V+K2V&4.3&0.135&89.6&1.052&0.854&1.018&0.83&5900&5000&13.0&11.1 \\
V432 Aur&F7V+F8V&3.1&0.0&90.0&1.22&1.08&2.46&1.23&6080&6685&50.0&20.0 \\
ER Vul&F9V+G0V&0.7&0.0&66.3&1.09&1.06&1.16&1.18&6000&5883&84.1&85.2 \\
CD Tau&F6V+F6V&3.4&0.0&87.7&1.442&1.368&1.798&1.584&6200&6200&28.0&26.0 \\
BH Vir&F8V+G5V&0.8&0.0&87.6&1.178&1.05&1.22&1.11&6000&5850&79.9&68.5 \\
HP Dra&F9V+F9V&10.8&0.037&87.8&1.133&1.094&1.371&1.052&6000&5895&4.2&4.4 \\
LV Her&F9V+F9V&18.4&0.613&89.6&1.193&1.17&1.358&1.313&6060&6030&13.0&13.0 \\
BF Dra&F6V+F6V&11.2&0.387&88.4&1.414&1.375&2.086&1.922&6360&6400&10.5&9.0 \\
BK Peg&F8&5.5&0.005&88.0&1.414&1.257&1.988&1.474&6265&6320&16.6&13.4 \\
2MASSJ05282082+0338327&K1V+K3V&3.9&0.0&83.7&1.375&1.329&1.83&1.73&5103&4751&24.6&24.6 \\
GX Gem&F7V+F7V&4.0&0.0&85.8&1.488&1.467&2.327&2.238&6194&6166&29.6&29.1 \\
BW Aqr&F7IV+F8IV&6.7&0.178&88.4&1.49&1.39&2.06&1.79&6350&6450&9.6&13.5 \\
RZ Cha&F5V+F5V&2.8&0.0&82.8&1.514&1.506&2.264&2.264&6457&6457&39.3&39.3 \\
RT And&F8V+K3+&0.6&0.012&88.4&1.24&0.91&1.26&0.9&6095&4732&100.0&80.0 \\
IT Cas&F6V+F6V&3.9&0.085&89.6&1.33&1.328&1.593&1.56&6470&6470&19.0&17.0 \\
V1130 Tau&F0&0.8&0.0&73.8&1.306&1.392&1.489&1.782&6650&6625&96.2&109.0 \\
DM Vir&F7V+F7V&4.7&0.0&90.0&1.454&1.448&1.763&1.763&6500&6500&20.0&20.0 \\
UX Men&F8V+F8V&4.2&0.003&89.6&1.238&1.198&1.348&1.274&6200&6150&16.4&15.1 \\
EI Cep&F3IV+F1V&8.4&0.006&87.2&1.772&1.68&2.896&2.329&6750&6950&13.0&17.0 \\
FS Mon&F2V+F4V&1.9&0.0&87.7&1.632&1.462&2.051&1.629&6715&6550&52.0&43.0 \\
HY Vir&F0m+F5V&2.7&0.0&81.6&1.838&1.404&2.806&1.519&7870&6546&48.5&23.2 \\
AD Boo&F6V+G0V&2.1&0.0&87.7&1.414&1.209&1.612&1.216&6575&6145&38.0&37.0 \\
RR Lyn&A6IV+F0V&9.9&0.079&87.4&1.927&1.507&2.57&1.59&7570&6980&14.6&11.3 \\
BP Vul&A7mV+F2mV&1.9&0.035&87.7&1.737&1.408&1.852&1.489&7709&6823&45.4&40.5 \\
XY Cet&A5V+A7V&2.8&0.0&87.7&1.773&1.615&1.873&1.773&7870&7620&34.4&34.1 \\
HD 71636&F2V+F5V&5.0&0.0&85.6&1.513&1.285&1.571&1.361&6950&6440&12.5&12.4 \\
RS Cha&A8V+A8V&1.7&0.0&83.2&1.89&1.87&2.15&2.36&7638&7228&64.5&70.5 \\
V505 Per&F5V+F5V&4.2&0.0&88.0&1.269&1.251&1.287&1.266&6512&6462&15.4&15.3 \\
SV Cam&F5V+K4V&0.6&0.0&73.8&1.47&0.87&1.38&0.94&6038&4804&120.8&82.3 \\
VZ Cep&F3V+G4V&1.2&0.0&80.0&1.402&1.108&1.534&1.042&6670&5720&57.9&50.8 \\
SZ Cen&A7V+A7V&4.1&0.0&88.1&2.317&2.277&4.552&3.62&8000&8280&60.0&44.0 \\
VZ Hya&F5V+F7V&2.9&0.0&88.9&1.271&1.146&1.314&1.112&6645&6290&21.0&20.0 \\
HS Hya&F4V+F4V&1.6&0.0&85.6&1.255&1.219&1.275&1.216&6500&6400&41.1&39.1 \\
V885 Cyg&A3Vm+A4mV&1.7&0.0&70.6&2.005&2.234&2.345&3.385&8375&8150&110.2&74.2 \\
KW Hya&A5Vm+F0-1V&7.8&0.095&87.6&1.978&1.488&2.126&1.484&8000&6900&15.0&13.0 \\
V570 Per&F3V+F5V&1.9&0.0&77.4&1.449&1.35&1.5&1.36&6842&6580&41.0&36.9 \\
V1031 Ori&A3V+A6V&3.4&0.0&85.6&2.473&2.287&4.321&2.987&7850&8400&22.1&43.1 \\
EW Ori&G0V+G5V&6.9&0.076&89.9&1.173&1.123&1.168&1.097&6070&5900&9.0&8.8 \\
SW CMa&A4-A5&10.1&0.317&88.6&2.239&2.104&3.014&2.495&8200&8100&24.0&10.0 \\
PV Pup&A0V+A2V&1.7&0.01&83.1&1.565&1.554&1.542&1.499&6920&6930&43.3&43.3 \\
EY Cep&F0+F0&8.0&0.443&89.9&1.523&1.498&1.463&1.468&7090&6970&10.0&10.0 \\
GZ CMa&A3Vm+A4:V&4.8&0.0&86.6&2.2&2.0&2.49&2.13&8810&8531&24.0&18.0 \\
WW Aur&A4m+A5m&2.5&0.0&87.6&1.964&1.814&1.927&1.841&7960&7670&35.0&37.0 \\
WW Cam&A4Vm+A4Vm&2.3&0.009&88.3&1.92&1.873&1.911&1.808&8360&8240&42.8&41.9 \\
V459 Cas&A1V+A1V&8.5&0.024&89.5&2.02&1.96&2.009&1.965&9140&9100&54.0&43.0 \\
HW CMa&A6&21.1&0.502&84.8&1.721&1.781&1.643&1.662&7560&7700&12.0&12.0 \\
VV Pyx&A1V+A1V&4.6&0.096&88.1&2.098&2.098&2.167&2.167&9500&9500&23.0&23.0 \\
V1229 Tau&A0V+Am&2.5&0.0&77.6&2.221&1.565&1.843&1.586&10025&7262&37.4&32.7 \\
PT Vel&A1V+A6V&1.8&0.127&88.2&2.198&1.626&2.094&1.559&9247&7638&63.0&40.0 \\
$\delta$ Vel&A1V&45.2&0.287&90.0&2.43&2.27&2.97&2.52&9450&9830&143.5&149.6 \\
CM Lac&A2V+A8V&1.6&0.0&86.8&1.98&1.5&1.51&1.55&9000&4586&44.1&34.1 \\
V364 Lac&A1Vm+A8Vm&7.4&0.287&89.2&2.333&2.296&3.307&2.985&8250&8500&45.0&15.0 \\
V396 Cas&A1V+A3V&5.5&0.0&88.8&2.398&1.901&2.592&1.779&9225&8550&16.0&21.0 \\
TV Nor&A+A&8.5&0.0&89.7&2.053&1.665&1.839&1.55&9120&7798&13.0&11.3 \\
YZ Cas&A2IV+F2V&4.5&0.0&88.3&2.31&1.35&2.53&1.35&9200&6700&34.0&16.0 \\
V821 Cas&A1.5V+A4V&1.8&0.138&82.7&2.04&1.62&2.31&1.39&9400&8450&70.6&57.5 \\
V526 Sgr&B9.5V+A2V&1.9&0.219&89.1&2.27&1.68&1.89&1.56&10140&8710&110.0&75.0 \\
CV Vel&B2.5V+ B2.5&6.9&0.0&86.6&6.066&5.972&4.126&3.908&18000&17780&19.0&31.1 \\
TZ Men&B9.5V+late A&8.6&0.035&88.7&2.487&1.504&2.016&1.432&10400&7200&16.0&12.0 \\
EE Peg&A3mV+F5V&2.6&0.0&88.6&2.156&1.335&2.089&1.312&8700&6450&40.0&26.0 \\
V392 Car&A2V+A2V&3.2&0.0&81.9&1.9&1.853&1.625&1.601&8850&8650&27.9&23.8 \\
V1647 Sgr&A1V+A2V&3.3&0.413&90.0&2.19&1.97&1.83&1.67&9600&9100&80.0&70.0 \\
CO And&F8V+F8V&3.7&0.0&86.9&1.289&1.264&1.727&1.694&6140&6170&24.1&23.8 \\
V906 Sco&B9V+B9V&2.8&0.005&76.2&3.378&3.253&4.521&3.515&10400&10700&82.4&63.9 \\
V731 Cep&B8.5V+A1.5V&6.1&0.017&88.7&2.577&2.017&1.823&1.717&10700&9265&19.0&18.0 \\
GG Ori&A2V+A2V&6.6&0.222&89.3&2.342&2.338&1.852&1.83&9950&9950&16.0&16.0 \\
V1143 Cyg&F5V+F5V&7.6&0.54&87.0&1.391&1.347&1.346&1.323&6460&6400&18.0&28.0 \\
$\chi^2$ Hya&B8V+B8III-IV&2.3&0.0&78.4&3.61&2.64&4.39&2.16&11750&11100&114.3&61.3 \\
AS Cam&B8V+B9.5V&3.4&0.135&88.8&3.3&2.5&2.6&1.98&12000&10700&40.0&30.0 \\
AR Aur&B9V+B9.6V&4.1&0.0&88.5&2.48&2.294&1.781&1.816&10950&10350&10.0&11.0 \\
IQ Per&B7.5V+A6V&1.7&0.075&89.3&3.51&1.73&2.45&1.5&12300&7670&68.0&44.0 \\
V413 Ser&B8V+B9V&2.3&0.193&79.9&3.68&3.36&3.21&2.93&11100&10350&44.7&38.6 \\
\hline
\end{tabular}
\end{table*}
\begin{table*}
\ContinuedFloat
\centering
\scriptsize
\caption{Continued.}
\label{tab:main}
\begin{tabular}{llrlrllllrrrr}
\hline
System & Type & $P$  & $e$ & $i$  &$M_1$ &$M_2$&$R_1$ &$R_2$ &$T_{\mathrm{eff},1}$ &$T_{\mathrm{eff},2}$& \veq$_{,1}$ & \veq$_{,2}$ \\ 
 & & [d] & & [$^\circ$] & [\Msun] &  [\Msun] & [$R_\odot$] & [$R_\odot$] & [K] & [K] & [\kms] & [\kms]\\ 
\hline
BD +03 3821&B8V&3.7&0.294&75.4&4.04&2.72&3.77&2.04&13140&12044&113.0&61.7 \\
$\zeta$ Phe&B6V+B8V&1.7&0.0&87.8&3.93&2.551&2.851&1.853&14550&11910&85.1&75.1 \\
MU Cas&B5V+B5V&9.7&0.193&87.0&4.66&4.57&4.19&3.67&14750&15100&21.0&22.0 \\
YY Sgr&B5V+B6V&2.6&0.158&88.9&3.9&3.48&2.56&2.33&14800&14125&58.0&39.0 \\
$\eta$ Mus&B8V+B8V&2.4&0.0&77.4&3.3&3.29&2.14&2.13&12700&12550&34.8&45.1 \\
EP Cru&B5V+B5V&11.1&0.187&89.7&5.02&4.83&3.59&3.495&15700&15400&141.4&137.8 \\
GG Lup&B7V+B9V&1.8&0.15&86.8&4.116&2.509&2.379&1.725&14750&11000&89.1&65.1 \\
U Oph&B5V+B5V&1.7&0.2&88.3&4.93&4.56&3.29&3.01&16900&16000&125.1&115.1 \\
V539 Ara&B3V+B4V&3.2&0.056&85.0&6.25&5.33&4.43&3.73&18200&16982&75.3&48.2 \\
IM Mon&B4V+B6.5V&1.2&0.0&62.2&5.5&3.32&3.15&2.36&17500&14500&166.2&101.7 \\
V1388 Ori&B2.5IV-V+ B3V&2.2&0.0&75.5&7.42&5.16&5.6&3.76&20500&18500&129.1&77.5 \\
V760 Sco&B4V+B4V&1.7&0.026&82.2&4.98&4.62&3.01&2.64&16900&16300&95.9&85.8 \\
LT CMa&B4V+B6.5V&1.8&0.059&74.2&5.59&3.36&3.59&2.04&17000&13140&113.3&69.6 \\
AG Per&B4V+B4V&2.0&0.071&81.4&5.36&4.9&2.99&2.6&18200&17400&95.1&70.8 \\
V615 Per&B7V&13.7&0.01&88.8&4.075&3.179&2.291&1.903&15000&11000&28.0&8.0 \\
V380 Cyg&B1.5 II-III+B2V&12.4&0.234&82.4&11.1&6.95&14.7&3.74&21350&20500&98.9&32.3 \\
V346 Cen&B1.5III+B2V&6.3&0.288&83.9&11.8&8.4&8.2&4.2&26500&24000&165.9&140.8 \\
SZ Cam&O9IV+B0.5V&2.7&0.0&75.2&14.31&10.69&8.91&6.7&30360&27244&150.0&123.1 \\
QX Car&B3V+B3V&4.5&0.278&85.7&9.266&8.48&4.29&4.05&23800&22600&120.3&110.3 \\
V453 Cyg&B0.4IV+B0.7IV&3.9&0.022&88.0&14.36&11.11&8.551&5.489&26600&25500&107.1&97.1 \\
AH Cep&B0.5V+B0.5V&1.8&0.0&69.0&15.4&13.6&6.38&5.86&29900&28600&198.2&198.2 \\
DW Car&B1V+B1V&1.3&0.0&85.7&11.34&10.63&4.558&4.297&27900&26500&182.5&177.5 \\
V578 Mon&B1V+early B&2.4&0.087&72.6&14.54&10.29&5.23&4.32&30000&26400&122.6&98.5 \\
HI Mon&B0V+B0.5V&1.6&0.367&80.0&14.2&12.2&5.13&4.99&30000&29000&152.3&152.3 \\
EM Car&O8V+O8V&3.4&0.012&81.5&22.89&21.42&9.35&8.34&34000&34000&131.4&151.7 \\
Y Cyg&O9.3V+O9.4V&3.0&0.154&86.7&17.5&17.3&6.0&5.7&31000&31570&138.2&130.2 \\
V3903 Sgr&O7V+O9V&1.7&0.0&65.2&27.27&19.01&8.088&6.125&38000&34100&253.4&187.3 \\
V451 Oph&B9V+B9V&2.2&0.013&85.9&2.776&2.356&2.64&2.028&10800&9800&41.1&30.1 \\
DH Cep&O6V+O7V&2.1&0.0&47.0&32.7&29.8&8.69&8.58&41000&39550&213.3&187.3 \\
\hline
\end{tabular}
\end{table*}

\subsection{Refined Theory for Stars with Convective Envelopes}
\label{sect:darwing_ce}
Stars with masses greater than $1.25~M_\odot$ possess outer envelopes in radiative equilibrium, while lower-mass stars exhibit convective envelopes (CE). For stars with surface convection and a radiative core, we follow \citet{zahn1977} to calculate the synchronisation timescale, $t_\sync^\mathrm{D}$, in years:
\begin{equation}
    t_\sync^\mathrm{D} = 0.4311 \frac{1}{q^2} \left( \frac{a}{R} \right)^6 \frac{1}{3k_2} \frac{I}{MR^2} \left( \frac{MR^2}{L} \right)^{1/3} \quad \text{yr},
    \label{eq:tsync_CE}
\end{equation}
where $M$, $R$, and $L$ represent the mass, radius, and luminosity of the star in solar units. The apsidal motion constant, $k_2$, is computed at each evolutionary step using the integral from the center to the surface of the star, as described by \citet{kopal1959}:
\begin{equation}
    k_2 = \frac{16\pi}{5MR^2} \int_0^R \rho(r) r^2 \, \mathrm{d}r,
\end{equation}
where $\rho(r)$ is the density at radius $r$. Tidal friction, driven by the variability of $\rho(r)$, is the primary mechanism for energy dissipation. 
Consequently, tidal synchronisation depends on the apsidal motion constant, $k_2$.

\subsection{Refined Theory for Stars with Radiative Envelopes}
\label{sect:darwing_re}
For stars featuring radiative envelopes (RE) and convective cores, the synchronisation timescale, $t_\sync^\mathrm{D}$, can be calculated using the expression provided by \citet{zahn1975} and discussed in \citet{hilditch2001}:
\begin{equation}
    t_\sync^\mathrm{D} = 3.1816 \times 10^{-6} \left( \frac{R^3}{M} \right)^{1/2} \frac{I}{MR^2} \frac{1}{E_2} \frac{1}{q^2} \left( \frac{a}{R} \right)^{8.5} \frac{1}{(1+q)^{5/6}} \quad \text{yr},
    \label{eq:tsync_RE}
\end{equation}
where $R$ and $M$ are the radius and mass of the star in solar units, and $E_2$ is the tidal torque constant, tabulated by \citet{claret2004}.

Equations~(\ref{eq:tsync_CE}) and (\ref{eq:tsync_RE}) refine Darwin's original theory, offering more precise synchronisation timescales for stars with convective and radiative envelopes, respectively.

\subsection{Tidal Action Produced by Meridional Circulation}
Meridional circulation naturally occurs in rotating stars and consists of large-scale circulation currents, emerging at the poles and descending into deeper layers at the equator. In binary systems, such circulation can be driven by the gravitational influence of a close companion \citep{hastings2020}. According to \citet{tassoul1992, tassoul2000} in binary stars, meridional circulation contributes an additional component to tidal interactions, characterized by the following timescale:
\begin{equation}
    t_\sync^\mathrm{T} = \frac{1.44 \times 10^{\sigma - N/4}}{q(1+q)^{3/8}} \left( \frac{L_\odot}{L} \right)^{1/4} \left( \frac{M_\odot}{M} \right)^{1/8} \left( \frac{R}{R_\odot} \right)^{9/8} \left( \frac{a}{R} \right)^{33/8} \quad \text{yr},
    \label{eq:tsyncT}
\end{equation}
where $\sigma$ and $N$ are the primary sources of uncertainty in Eq.~(\ref{eq:tsyncT}). 
The superscript "T" refers to Tassoul.
The parameter $\sigma$ is a dimensionless factor that parameterizes the overall efficiency of the large-scale meridional flow. 
It acts as an inverse measure of the tidal coupling strength: smaller values correspond to more efficient dissipation and stronger tidal friction (e.g., $\sigma=3.5$), while larger values  (e.g., $\sigma=5$) indicate weaker tidal interactions. 
Choosing $\sigma=0$ in Eq.~(\ref{eq:tsyncT}) yields the e-folding time, which is a lower limit for the real synchronization time scale \citep{claret1995}. The left panel of Figure~\ref{fig:sync_factor_vs_age} suggests that $\sigma=4$ is a reasonable choice for the binaries in this paper containing a large fraction of synchronized systems.
The parameter $N$ is defined as a function of the turbulent and radiative viscosity coefficients:
\begin{equation}
    N = \log{\left(1 + \frac{\nu_\mathrm{turb}}{\nu_\mathrm{rad}}\right)}.
    \label{eq:N}
\end{equation}
For a radiative envelope without turbulence, $N = 0$. 
In early-type binaries, \citet{tassoul1988} inferred a value of 10 from observations  of systems in the Hyades, Praesepe and M67 clusters.  We therefore adopt $N = 10$ for convective envelopes, as turbulent viscosity dominates over radiative viscosity in convective regions. 

\begin{figure}[h]
    \includegraphics[width=\linewidth]{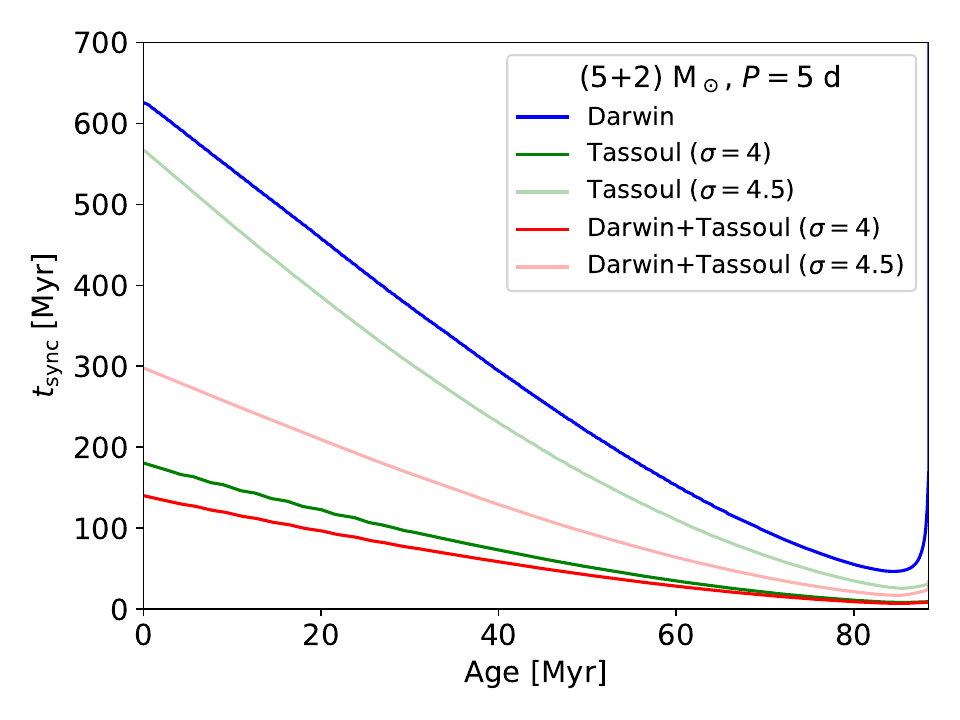}
    \caption{Evolution of the synchronisation times for the primary of the same system (Darwin mode in blue, Tassoul mode in green, and the mode where both contributions act together in red).}
    \label{fig:timescales}
\end{figure}

In this work, both the Darwin and Tassoul mechanisms operate simultaneously. The $(a/R)$-dependence of the meridional circulation contribution to the synchronisation timescale is given by:
\begin{equation}
    t_\sync^\mathrm{T} \propto \left( \frac{a}{R} \right)^{4.125}.
\end{equation}

For comparison, the Darwin contribution for a star with a convective envelope (CE) is:
\begin{equation}
    t_\sync^\mathrm{D}(\mathrm{CE}) \propto \left( \frac{a}{R} \right)^6 \quad \text{yr},
\end{equation}
and for a star with a radiative envelope (RE):
\begin{equation}
    t_\sync^\mathrm{D}(\mathrm{RE}) \propto \left( \frac{a}{R} \right)^{8.5} \quad \text{yr}.
\end{equation}
The contribution of meridional circulation to synchronisation becomes particularly significant at larger separations. The total synchronisation timescale accounting for both effects is given by:
\begin{equation}
    \frac{1}{t_\sync} = \frac{1}{t_\sync^\mathrm{D}} + \frac{1}{t_\sync^\mathrm{T}}.
\end{equation}
An illustrative example of these timescales is given in Fig.~\ref{fig:timescales}: the resulting synchronisation timescale (red) is the combination of the Darwin (blue) and Tassoul (green) timescales.  To illustrate the sensitivity of the $\sigma$ parameter, we show the timescales for $\sigma=4$ and $\sigma=4.5$. In the rest of this work, we adopt $\sigma=4$. Further numerical details can be found in \citet{vanrensbergen2016}.

\begin{figure*}
    \centering
    \includegraphics[width=0.49\linewidth]{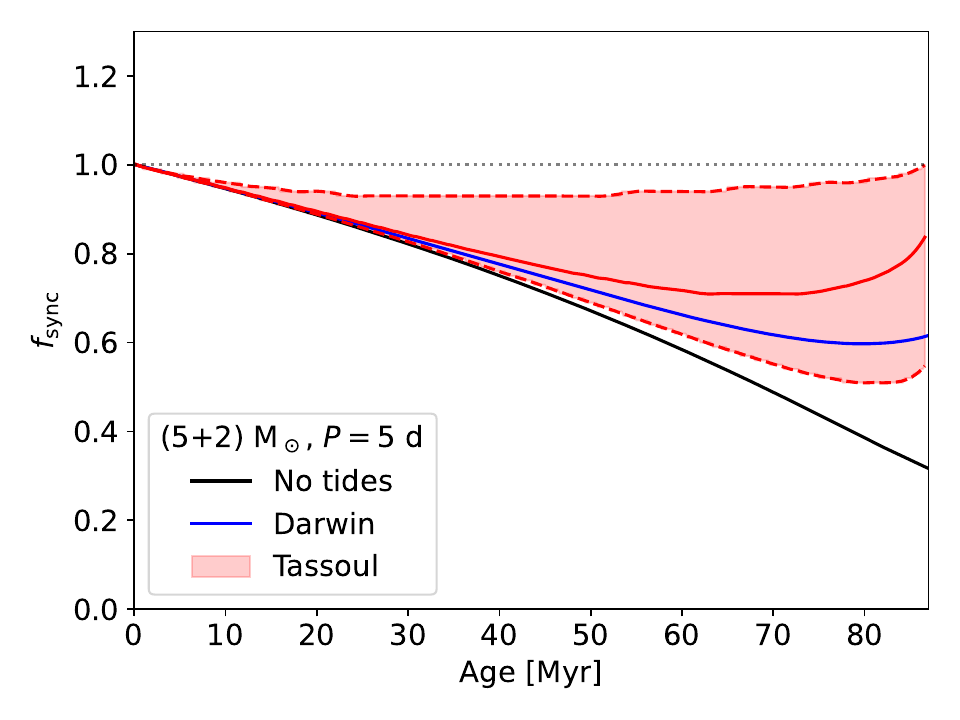}
    \includegraphics[width=0.49\linewidth]{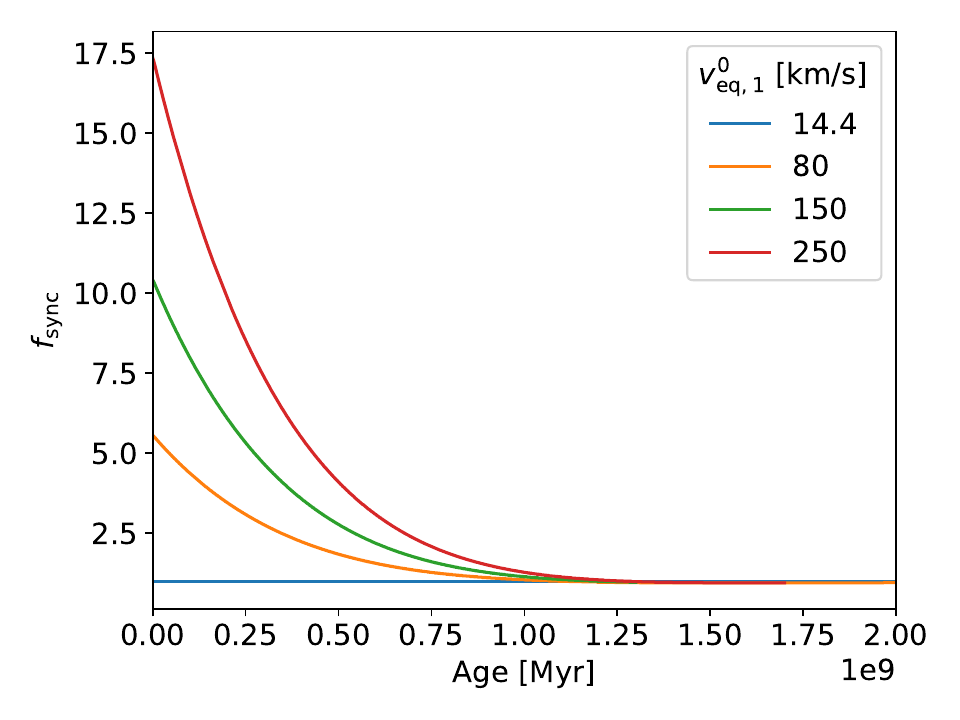}
    \caption{Left: Evolution of the synchronisation factor $f_\sync = v_\mathrm{eq}/v_\mathrm{eq}^\sync$ for the primary of a (5+2)~M$_\odot$ binary system with an initial orbital period of 5~d, when no tidal interactions are considered (black), when Darwin theory is applied (blue) and when meridional circulation is added (red shaded area). The lower dashed red border is the weak version ($\sigma=5$) and the upper one is the strong version ($\sigma=3$) of Eq.~\ref{eq:tsyncT}. The plain red line is the version with $\sigma=4$, adopted in this work. The evolution of this system covers the main-sequence. Right: Evolution of the synchronisation factor for primary in DM~Vir  ($\sigma=4$). Because the system is synhronised today, no unique ZAMS spin lead to the current value of equatorial velocity.}
    \label{fig:sync_factor_vs_age}
\end{figure*}

\subsection{Synchronisation factor}
The synchronisation factor $f_\sync$ for each component is defined as the ratio of the equatorial velocity $v_\mathrm{eq}$ to the synchronisation one $v_\mathrm{eq}^\sync$, $f_\sync = v_\mathrm{eq}/v_\mathrm{eq}^\sync$.  
When the rotational period of a component is higher than the orbital period of the system ($f_\sync >1$),  which is the most frequent situation, tidal effects tend to reduce $f_\sync$ towards unity over time and synchronise the rotational period with orbital period.
If $f_\sync<1$, tidal effects will increase the rotational period to reach synchronisation, \emph{i.e.} $f_\sync=1$.
Left panel of Fig.~\ref{fig:sync_factor_vs_age} illustrates the evolution of the synchronisation factor for the primary star in a (5+2) $M_\odot$ binary system in a circular orbit. In this example, the system starts on the ZAMS with synchronous rotation and orbital periods of 5 days. During the main sequence, the expansion of the star produces a spin down which is seen by the decrease of $f_\sync$ (black line). When tidal interactions are included using the Darwin formalism (blue line), synchronisation occurs more rapidly in closer systems. Incorporating meridional circulation (Tassoul formalism) can significantly shorten the synchronisation timescale. 

\section{Angular Momentum Loss via Stellar Winds}
Mass lost through stellar winds carries away angular momentum not only from the individual stellar components (resulting in $\Delta J_{\mathrm{wind}} < 0$) but also from the orbit. In this study, we calculate the rate of orbital angular momentum loss, $\dot{J}_{\mathrm{wind}}$, under the Jeans mode approximation for a fast isotropic wind. The expression is given by:
\begin{equation}
\dot{J}_{\mathrm{wind}} = \dot{M}_{\mathrm{wind}} \omega_{\mathrm{orb}} r^2
\end{equation}
where $r$ is the distance from the star to the centre of mass of the system. 
Assuming stellar wind isotropy, we neglect any loss of rotational angular momentum from the individual stellar components. For the mass-loss rates, $\dot{M}_\wind$, we use the formulation of \citet{vink2001} for stars with effective temperatures, $\Teff > 12500$ K, and from \citet{dejager1988} for stars within the ranges $3.3 \leq \log{\Teff} \leq 4.8$ and $2.5 \leq \log{L/L_\odot} \leq 6.7$.
However, a significant fraction of the stars in our sample have $\log{\Teff}$ and $\log{L/L_\odot}$ values below these thresholds. For these stars, we use a linear approximation for the mass-loss rate provided by \citet{dejager1988}:
\begin{equation}
    \log{\dot{M}_\wind} = 1.769 \log{\left(\frac{L}{L_\odot}\right)} - 1.676 \log{\Teff} - 8.158.
\end{equation}

To align with the observed solar wind mass-loss rate, we adjusted the constant term in the equation to $-7.394$ for the cooler and fainter stars in our sample. This adjustment ensures that the derived mass-loss rates are consistent with the solar wind value, although the exact value of the constant has a negligible impact on the derived stellar ages.

\begin{table*}
\centering
\caption{Hydrogen mass fractions and ages calculated for 108 detached eclipsing binary systems, by decreasing ages.}
\scriptsize
\begin{tabular}{lcrrl|lcrrl}
\hline
\hline 
System & Spectral type & $X_\mathrm{C,1}$ &	$X_\mathrm{C, 2}$ &	Age [yr] & System & Spectral type & $X_\mathrm{C,1}$ &	$X_\mathrm{C, 2}$ &	Age [yr] \\	
\hline
UV Psc                &G5V+K3V   & 0.011&0.434 & $(8.64\pm2.17)\times10^9$    & V1229 Tau  &A0V+Am         & 0.349&0.583 & $(5.03\pm2.42)\times10^8$ \\
GZ Leo                &K1V+K1V   & 0.447&0.456 & $(7.41\pm1.29)\times10^9$    & PT Vel     &A1V+A6V        & 0.380&0.577 & $(4.81\pm0.84)\times10^8$ \\
CG Cyg                &G9V+K3V   & 0.315&0.477 & $(6.13\pm2.82)\times10^9$    & $\delta$ Vel&A1V 		   & 0.229&0.352 & $(4.66\pm0.13)\times10^8$ \\
V565 Lyr              &G+G       & 0.223&0.313 & $(5.96\pm0.61)\times10^9$    & CM Lac     &A2V+A8V        & 0.492&0.603 & $(4.63\pm3.53)\times10^8$ \\
V1430 Aql             &G2V+K4V   & 0.365&0.505 & $(4.96\pm2.24)\times10^9$    & V364 Lac   &A1Vm+A8Vm      & 0.315&0.524 & $(4.55\pm0.27)\times10^8$ \\
NGC188 KR V12         &G+G       & 0.018&0.093 & $(4.79\pm0.15)\times10^9$    & V396 Cas   &A1V+A3V        & 0.315&0.524 & $(4.55\pm0.27)\times10^8$ \\
AI Phe                &F7V+K0IV  & 0.000&0.000 & $(4.27\pm0.22)\times10^9$    & TV Nor     &A+A            & 0.486&0.585 & $(4.29\pm0.55)\times10^8$ \\
V636 Cen              &G1V+K2V   & 0.291&0.513 & $(4.15\pm2.58)\times10^9$    & YZ Cas     &A2IV+F2V       & 0.392&0.633 & $(4.11\pm0.43)\times10^8$ \\
V432 Aur              &F7V+F8V   & 0.000&0.268 & $(3.89\pm0.36)\times10^9$    & V821 Cas   &A1.5V+A4V      & 0.522&0.607 & $(3.75\pm2.16)\times10^8$ \\
ER Vul                &F9V+G0V   & 0.338&0.383 & $(3.22\pm0.60)\times10^9$    & V526 Sgr   &B9.5V+A2V      & 0.452&0.600 & $(3.68\pm1.11)\times10^8$ \\
CD Tau                &F6V+F6V   & 0.280&0.373 & $(2.85\pm0.02)\times10^9$    & CV Vel     &B2.5V+B2.5     & 0.400&0.416 & $(3.67\pm0.07)\times10^8$ \\
BH Vir                &F8V+G5V   & 0.351&0.497 & $(2.21\pm1.02)\times10^9$    & TZ Men     &B9.5V+late A   & 0.394&0.630 & $(3.39\pm1.27)\times10^8$ \\
HP Dra                &F9V+F9V   & 0.425&0.465 & $(2.16\pm1.28)\times10^9$    & EE Peg     &A3mV+F5V       & 0.524&0.649 & $(3.34\pm1.03)\times10^8$ \\
LV Her                &F9V+F9V   & 0.339&0.379 & $(2.14\pm0.04)\times10^9$    & V392 Car   &A2V+A2V        & 0.582&0.590 & $(3.17\pm0.09)\times10^8$ \\
BF Dra                &F6V+F6V   & 0.243&0.302 & $(2.01\pm0.01)\times10^9$    & V1647 Sgr  &A1V+A2V        & 0.539&0.583 & $(2.89\pm0.18)\times10^8$ \\
BK Peg                &F8        & 0.255&0.461 & $(1.97\pm0.10)\times10^9$    & CO And     &F8V+F8V        & 0.293&0.358 & $(2.68\pm1.80)\times10^8$ \\
2MASSJ05282082+0338327&K1V+K3V   & 0.327&0.382 & $(1.92\pm0.08)\times10^9$    & V906 Sco   &B9V+B9V        & 0.125&0.221 & $(2.16\pm0.09)\times10^8$ \\
GX Gem                &F7V+F7V   & 0.182&0.217 & $(1.87\pm0.02)\times10^9$    & V731 Cep   &B8.5V+A1.5V    & 0.527&0.617 & $(2.00\pm0.87)\times10^8$ \\
BW Aqr                &F7IV+F8IV & 0.257&0.382 & $(1.70\pm0.21)\times10^9$    & GG Ori     &A2V+A2V        & 0.575&0.576 & $(1.97\pm0.05)\times10^8$ \\
RZ Cha                &F5V+F5V   & 0.218&0.229 & $(1.69\pm0.02)\times10^9$    & V1143 Cyg  &F5V+F5V        & 0.665&0.669 & $(1.97\pm0.23)\times10^8$ \\
RT And                &F8V+K3+   & 0.362&0.614 & $(1.68\pm1.09)\times10^9$    & $\chi^2$ Hya&B8V+B8III-IV  & 0.075&0.523 & $(1.91\pm0.10)\times10^8$ \\
IT Cas                &F6V+F6V   & 0.443&0.445 & $(1.61\pm0.05)\times10^9$    & AS Cam     &B8V+B9.5V      & 0.388&0.581 & $(1.58\pm0.35)\times10^8$ \\
V1130 Tau             &F0        & 0.389&0.442 & $(1.53\pm0.12)\times10^9$    & AR Aur     &B9V+B9.6V      & 0.585&0.609 & $(1.58\pm0.47)\times10^8$ \\
DM Vir                &F7V+F7V   & 0.401&0.405 & $(1.35\pm0.01)\times10^9$    & IQ Per     &B7.5V+A6V      & 0.365&0.660 & $(1.45\pm0.59)\times10^8$ \\
UX Men                &F8V+F8V   & 0.452&0.491 & $(1.33\pm0.01)\times10^9$    & V413 Ser   &B8V+B9V        & 0.316&0.430 & $(1.40\pm0.13)\times10^8$ \\
EI Cep                &F3IV+F1V  & 0.144&0.265 & $(1.19\pm0.01)\times10^9$    & BD +03 3821&B8V            & 0.207&0.580 & $(1.29\pm0.14)\times10^8$ \\
FS Mon                &F2V+F4V   & 0.348&0.467 & $(1.12\pm0.02)\times10^9$    & $\zeta$ Phe&B6V+B8V        & 0.358&0.616 & $(1.11e\pm0.7)\times10^8$ \\
HY Vir                &F0m+F5V   & 0.199&0.503 & $(1.08\pm0.02)\times10^9$    & MU Cas     &B5V+B5V        & 0.333&0.361 & $(7.54\pm0.32)\times10^7$ \\
AD Boo                &F6V+G0V   & 0.526&0.559 & $(8.98\pm2.72)\times10^8$    & YY Sgr     &B5V+B6V        & 0.535&0.583 & $(6.78\pm0.56)\times10^7$ \\
RR Lyn                &A6IV+F0V  & 0.233&0.504 & $(8.70\pm0.14)\times10^8$    & $\eta$ Mus &B8V+B8V        & 0.609&0.610 & $(6.20\pm0.04)\times10^7$ \\
BP Vul                &A7mV+F2mV & 0.412&0.544 & $(8.24\pm0.80)\times10^8$    & EP Cru     &B5V+B5V        & 0.402&0.445 & $(5.58\pm0.35)\times10^7$ \\
XY Cet                &A5V+A7V   & 0.406&0.480 & $(8.06\pm1.03)\times10^8$    & GG Lup     &B7V+B9V        & 0.484&0.612 & $(5.27\pm2.08)\times10^7$ \\
HD 71636              &F2V+F5V   & 0.522&0.581 & $(7.91\pm0.88)\times10^8$    & U Oph      &B5V+B5V        & 0.456&0.511 & $(5.10\pm0.38)\times10^7$ \\
RS Cha                &A8V+A8V   & 0.336&0.345 & $(7.86\pm0.42)\times10^8$    & V539 Ara   &B3V+B4V        & 0.271&0.457 & $(4.31\pm0.51)\times10^7$ \\
V505 Per              &F5V+F5V   & 0.617&0.659 & $(7.28\pm1.30)\times10^8$    & IM Mon     &B4V+B6.5V      & 0.471&0.646 & $(3.78\pm1.02)\times10^7$ \\
SV Cam                &F5V+K4V   & 0.552&0.668 & $(7.18\pm4.12)\times10^8$    & V1388 Ori  &B2.5IV-V+B3V   & 0.203&0.532 & $(3.60\pm0.27)\times10^7$ \\
VZ Cep                &F3V+G4V   & 0.571&0.624 & $(7.01\pm4.00)\times10^8$    & V760 Sco   &B4V+B4V        & 0.556&0.585 & $(3.42\pm0.36)\times10^7$ \\
SZ Cen                &A7V+A7V   & 0.000&0.040 & $(6.81\pm0.01)\times10^8$    & LT CMa     &B4V+B6.5V      & 0.475&0.647 & $(2.64\pm0.19)\times10^7$ \\
VZ Hya                &F5V+F7V   & 0.617&0.616 & $(6.79\pm1.06)\times10^8$    & AG Per     &B4V+B4V        & 0.593&0.616 & $(2.24\pm0.40)\times10^7$ \\
HS Hya                &F4V+F4V   & 0.666&0.595 & $(6.59\pm2.06)\times10^8$    & V615 Per   &B7V            & 0.653&0.675 & $(2.07\pm0.15)\times10^7$ \\
V885 Cyg              &A3Vm+A4mV & 0.120&0.360 & $(6.41\pm0.23)\times10^8$    & V380 Cyg   &B1.5II-III+B2V & 0.000&0.514 & $(1.94\pm0.01)\times10^7$ \\
KW Hya                &A5Vm+F0-1V& 0.377&0.565 & $(6.40\pm0.28)\times10^8$    & V346 Cen   &B1.5III+B2V    & 0.224&0.502 & $(1.42\pm0.03)\times10^7$ \\
V570 Per              &F3V+F5V   & 0.575&0.594 & $(6.35\pm2.12)\times10^8$    & SZ Cam     &O9IV+B0.5V     & 0.135&0.456 & $(1.18\pm0.13)\times10^7$ \\
V1031 Ori             &A3V+A6V   & 0.049&0.190 & $(6.30\pm0.29)\times10^8$    & QX Car     &B3V+B3V        & 0.517&0.551 & $(1.12\pm0.08)\times10^7$ \\
EW Ori                &G0V+G5V   & 0.617&0.630 & $(6.15\pm0.75)\times10^8$    & V453 Cyg   &B0.4IV+B0.7IV  & 0.232&0.441 & $(1.07\pm0.02)\times10^7$ \\
SW CMa                &A4-A5     & 0.190&0.316 & $(6.10\pm0.08)\times10^8$    & AH Cep     &B0.5V+B0.5V    & 0.431&0.493 & $(6.21\pm0.25)\times10^6$ \\
PV Pup                &A0V+A2V   & 0.555&0.557 & $(6.08\pm0.75)\times10^8$    & DW Car     &B1V+B1V        & 0.575&0.590 & $(5.79\pm0.01)\times10^6$ \\
EY Cep                &F0+F0     & 0.576&0.581 & $(5.65\pm0.35)\times10^8$    & V578 Mon   &B1V+early B    & 0.507&0.603 & $(5.59\pm1.50)\times10^6$ \\
GZ CMa                &A3Vm+A4:V & 0.294&0.422 & $(5.61\pm0.67)\times10^8$    & HI Mon     &B0V+B0.5V      & 0.532&0.575 & $(5.12\pm1.15)\times10^6$ \\
WW Aur                &A4m+A5m   & 0.435&0.495 & $(5.57\pm0.47)\times10^8$    & EM Car     &O8V+O8V        & 0.368&0.399 & $(4.81\pm0.08)\times10^6$ \\
WW Cam     			  &A4Vm+A4Vm & 0.473&0.489 & $(5.37\pm0.26)\times10^8$    & Y Cyg      &O9.3V+O9.4V    & 0.556&0.558 & $(3.43\pm0.30)\times10^6$ \\
V459 Cas   			  &A1V+A1V   & 0.433&0.460 & $(5.30\pm0.19)\times10^8$    & V3903 Sgr  &O7V+O9V        & 0.495&0.569 & $(2.85\pm0.12)\times10^6$ \\
HW CMa     			  &A6        & 0.530&0.546 & $(5.14\pm0.30)\times10^8$    & V451 Oph   &B9V+B9V        & 0.387&0.522 & $(2.63\pm0.10)\times10^6$ \\
VV Pyx                &A1V+A1V   & 0.409&0.409 & $(5.11\pm0.01)\times10^8$    & DH Cep     &O6V+O7V        & 0.495&0.517 & $(2.38\pm0.22)\times10^6$ \\
\hline
\end{tabular}
\tablefoot{ Indices 1 and 2 designed primary and secondary components, respectively.}
\label{tab:age}
\end{table*}

\section{Calculations and Results}
We calculated the evolution of the binary systems using the Brussels Binary Evolution code, which incorporates tidal interactions and angular momentum loss through stellar winds, as described in the preceding sections. This code is an extended version of the original code of \citet{paczynksi1971} and further developed by \citet{de_loore1975}. The code models binary stars assuming circular orbits. As a result, the impact of orbital eccentricity on the inferred ages and zero‑age main‑sequence (ZAMS) rotation rates is not accounted for.
In our sample, there are $29/108=27$\% of systems with $e>0.1$. Among the non-synchronized systems this increases to $12/29=41$\% with $e>0.1$. Therefore, the derived ages and ZAMS spins for these eccentric systems should be treated with caution.

\subsection{Determination of the Current Ages of Binaries}
The 108 binaries, detailed in Section~\ref{sect:sample} and Table~\ref{tab:main} were  analyzed in this study. The calculations begin at the ZAMS and although both components of these binaries are still on the main sequence, we extended our calculations beyond their current observed states.

The ages of the binary components were determined by identifying the time at which the calculated radii of both stars best match the measured radii reported in the catalogue of \citet{eker2014}. 
Indeed, stellar radii and masses in eclipsing binaries are determined purely from orbital dynamics and eclipse geometry with a precision often better than 3\% while luminosities and effective temperatures are generally subject to larger uncertainties from bolometric corrections and atmospheric models. The monotonic increase in the stellar radius on the main sequence and up to the RGB serves as a robust indicator to estimate the age.
Since there is often a time difference between the age estimate when the calculated radii match the measured radii of each component, we used the mean of these times to determine the system's age. The age uncertainties are estimated as half of the age interval obtained from matching the radius of each stellar component.
This approach results in a mean relative age uncertainties of about 20\%, reaching 50\% for systems such as SV~Cam and $\zeta$~Phe.
 
Table~\ref{tab:age} presents the derived ages of the 108 binaries. The table lists the estimated age of each system.
Table~\ref{tab:age} also includes the values of $X_\mathrm{c,1}$ and $X_\mathrm{c,2}$, representing the mass fraction of hydrogen in the core of the stars. Values close to 0.7 indicate a young binary system, while values close to 0.0 suggest that the binary is nearing hydrogen exhaustion in the core. In a few cases where $X_c = 0$, the star has already left the main sequence on the HRD. Even in these cases, the age of the system can still be calculated.

\begin{table*}
\centering
\caption{Spins and synchronisation ratios computed at birth (ZAMS) and calculated nowadays for 29 systems, by decreasing $f_\mathrm{sync, 1}^0$. 
}
\scriptsize
\label{tab:sync}
\begin{tabular}{lrrrrrrrrrrrrrr}
\hline
\hline
&& \multicolumn{8}{c}{ZAMS} && \multicolumn{4}{c}{Observed today} \\
\cline{3-10} \cline{12-15} 
System & Spectral type & $R_1$ & $R_2$ & $T_\mathrm{eff,1}$ & $T_\mathrm{eff,2}$ & \veq$_{,1}^0$& \veq$_{,2}^0$ & $f_\mathrm{sync, 1}^0$&   $f_\mathrm{sync, 2}^0$&& \veq$_{,1}$& \veq$_{,2}$  & $f_\mathrm{sync, 1}$ &  $f_\mathrm{sync, 2}$\\
$\delta$ Vel$^*$&A1V & 1.594&1.539&11194&10641 & 171.00&166.00&95.70&96.22 && 143.22&149.38&41.70&53.95 \\
V364 Lac$^*$&A1Vm+A8Vm & 1.56&1.548&10864&10739 & 320.00&6.00&29.81&0.56 && 45.08&15.02&2.02&0.72 \\
EP Cru$^*$&B5V+B5V & 2.401&2.348&17988&17538 & 168.20&160.40&15.02&14.63 && 141.50&137.78&8.08&8.74 \\
YZ Cas&A2IV+F2V & 1.553&1.305&12105&6668 & 230.25&19.40&13.09&1.31 && 34.25&16.01&1.28&1.04 \\
V346 Cen$^*$&B1.5III+B2V & 3.906&3.223&28575&24043 & 354.00&217.50&11.19&8.49 && 165.45&140.60&2.59&4.09 \\
QX Car$^*$&B3V+B3V & 3.408&3.24&25351&24210 & 429.00&316.00&11.14&8.63 && 120.55&110.39&2.43&2.46 \\
V459 Cas&A1V+A1V & 1.456&1.437&9749&9527 & 96.00&71.25&11.02&8.29 && 54.00&42.93&4.41&3.71 \\
V380 Cyg$^*$&B1.5 II-III+B2V & 3.775&2.894&27733&21727 & 149.00&34.00&9.69&2.88 && 98.63&32.38&1.63&2.10 \\
SW CMa$^*$&A4-A5 & 1.582&1.483&10543&10069 & 49.50&11.50&6.49&1.55 && 24.10&10.08&1.57&0.82 \\
V392 Car&A2V+A2V & 1.419&1.405&9289&9099 & 140.00&21.50&6.19&0.96 && 27.61&24.58&1.08&0.96 \\
EI Cep&F3IV+F1V & 1.384&1.362&8770&8356 & 46.40&42.60&5.59&5.21 && 13.18&16.97&0.77&1.22 \\
VV Pyx&A1V+A1V & 1.481&1.481&10046&10046 & 80.00&89.00&4.90&4.90 && 23.03&23.08&0.96&0.96 \\
BK Peg&F8 & 1.352&1.218&6998&6338 & 52.00&14.20&4.26&1.26 && 16.60&13.40&0.87&1.00 \\
HW CMa$^*$&A6 & 1.386&1.371&8810&8531 & 12.40&12.40&3.73&3.77 && 11.96&12.10&2.99&3.10 \\
V615 Per&B7V & 2.129&1.847&15812&10069 & 28.50&8.10&3.63&1.19 && 28.13&8.05&3.35&1.14 \\
BF Dra$^*$&F6V+F6V & 1.325&1.316&6998&6792 & 20.50&14.50&3.43&2.44 && 10.50&9.10&1.11&1.05 \\
RR Lyn&A6IV+F0V & 1.427&1.335&9397&7498 & 23.50&13.50&3.24&1.99 && 14.45&11.43&1.10&1.42 \\
V731 Cep&B8.5V+A1.5V & 1.644&1.455&11668&9749 & 22.90&21.10&2.92&2.47 && 18.99&17.98&1.09&1.34 \\
KW Hya&A5Vm+F0-1V & 1.443&1.332&9594&7396 & 21.00&15.00&2.23&1.36 && 14.91&13.11&1.05&1.36 \\
V1143 Cyg$^*$&F5V+F5V & 1.32&1.303&6870&6668 & 19.26&30.65&2.20&3.55 && 18.05&28.05&2.03&3.19 \\
TV Nor&A+A & 1.467&1.359&9885&8279 & 18.95&12.80&2.18&1.59 && 13.02&11.34&1.14&1.26 \\
TZ Men&B9.5V+late A & 1.613&1.334&11376&7481 & 20.00&12.90&2.10&1.64 && 16.06&12.09&1.10&1.45 \\
MU Cas$^*$&B5V+B5V & 2.3&2.274&17179&16982 & 25.00&26.50&2.07&2.26 && 20.91&22.12&1.00&1.11 \\
GG Ori$^*$&A2V+A2V & 1.564&1.562&10889&10889 & 18.00&18.00&1.51&1.51 && 16.04&16.04&1.14&1.14 \\
EY Cep$^*$&F0+F0 & 1.336&1.334&7585&7447 & 10.50&10.50&1.24&1.24 && 10.09&10.09&1.08&1.09 \\
V505 Per&F5V+F5V & 1.232&1.211&6382&6324 & 13.80&16.20&0.93&1.12 && 15.45&15.22&0.99&1.01 \\
CV Vel&B2.5V+ B2.5 & 2.677&2.653&20090&19906 & 15.91&37.45&0.81&1.92 && 18.79&31.40&0.63&1.08 \\
HD 71636&F2V+F5V & 1.335&1.25&7533&6426 & $-6.50$&5.50&$-0.48$&0.44 && 12.55&12.44&0.80&0.91 \\
V396 Cas&A1V+A3V & 1.583&1.419&11091&9289 & $-77.00$&42.00&$-5.29$&3.22 && 15.98&20.96&0.63&1.31 \\
\hline
\end{tabular}
\tablefoot{Superscript 0 is for ZAMS. Indices 1 and 2 design primary and secondary components, respectively. Systems with an $^*$ have an eccentricity larger than 0.1. Negative velocities at ZAMS are retrograde with respect tot he orbit.}
\end{table*}

\subsection{Determination of the Initial Stellar Spins}
\label{sec:spin}
For each binary system listed in Table~\ref{tab:age}, we randomly selected initial equatorial velocities at the ZAMS for both stars. Using the ages determined in the previous section, we calculated the current equatorial velocities. This process was iterated until the calculated equatorial velocities matched the observed values with a precision better than 1\%. We noted that the initial equatorial velocities do not significantly affect the derived ages because the rotational kinetic energy of the orbit is several orders of magnitude larger than the roationatal kinetic energy of the stellar components.
Among the 108 binaries, many systems are currently synchronized. 
For them, we cannot determined a unique ZAMS rotational velocity because many different values lead to synchronisation, as illustrated for the system DM~Vir (right panel of Fig.~\ref{fig:sync_factor_vs_age}).
The estimation of initial spin velocities is therefore limited to the 29 binaries that contain at least one non‑synchronized component.

We computed the synchronization factors, defined as the ratio of the equatorial velocity to the synchronized velocity, $f_{\text{sync}} = v_{\mathrm{eq}}/v_\mathrm{eq}^\sync$, at both the ZAMS ($f_{\text{sync}}^0$) and the present day ($f_{\text{sync}}$). We computed the initial velocities $v_{\mathrm{eq}}^0$ from our evolutionary models and the current $v_{\mathrm{eq}}$ values derived from the observed $v \sin i$, adopting the orbital inclinations from \citet{eker2014} under the assumption of spin-orbit alignment. The present-day velocities are well reproduced by our models within a 1\% precision (see Table~\ref{tab:sync}). Notably, even the largest initial synchronization factors $f_{\text{sync}}^0$ result in equatorial velocities that remain well below the critical (break-up) velocity $v_{\mathrm{crit}}$. It is important to specify that for stars in binary systems, $v_{\mathrm{crit}}$ is lower than for isolated stars due to the reduction in effective surface gravity caused by tidal distortion and the proximity of the Roche lobe \citep{Van_Hamme1990}. While this effect is most pronounced as a star approaches RLOF, we account for it here to ensure that our initial model assumptions for $v_{\mathrm{eq}}^0$ do not lead to premature mass loss or non-physical rotation rates during the early evolution of the shorter-period systems in our sample (see Sect.~\ref{sect:sample}).

Two systems have very high initial synchronisation factors and are currently far from synchronized: $\delta$~Vel and V565~Lyr, with $f_\sync^0$ values of 95.7 and 84.6 for the primaries, and 96.2 and 85.2 for the secondaries, respectively. According to Table~\ref{tab:age}, $\delta$~Vel has not reached synchronisation at the age of 0.466 Gyr, contrarily to V565 Lyr at an age of 6.13 Gyr.

For all systems except HD~71636 and V396~CMa, the synchronisation factors at the ZAMS are larger than those measured presently, indicating that tidal interactions have slowed their rotation over time. Table~\ref{tab:sync} shows the radii, effective temperatures, and equatorial velocities computed at the ZAMS for the 29 systems, along with the present-day equatorial velocities and synchronisation factors. The corresponding observed radii and effective temperatures from \citet{eker2014} can be found in Table~\ref{tab:main}.

Retrograde spin in binary systems, where a component's rotation is opposite to its orbital motion, is a known theoretical possibility, particularly in systems formed via tidal capture \citep{lai1997}. While tidal interactions generally drive systems toward alignment and prograde synchronization over time, a retrograde signature can persist if the initial angular momentum was strongly misaligned and the system has not yet reached tidal equilibrium. In our modeling of \citet{eker2014} sample, we determined the ZAMS equatorial velocities, $v_{\mathrm{eq}}^0$, required to match the presently observed values within a 1\% precision. 
However, for the primary components of HD~71636 and V396~Cas, no assumption of an initially prograde rotation yields present-day velocities consistent with the observations. We find instead that the observed prograde velocities (12.55~km/s and 15.98~km/s, respectively) can only be reproduced if the stars began their main-sequence evolution with retrograde rotation ($-6.50$ and $-77.00$~km/s from Table~\ref{tab:sync}).
Although these stars have since evolved to exhibit prograde rotation due to tidal torques, their initial retrograde state is necessary to explain their current rotation rates. This is not without physical precedent; for instance, \citet{hale1994} found that approximately 30\% of solar-type binaries exhibit significant spin-orbit misalignment, and within our own solar system, Venus serves as a classic example of retrograde rotation in a prograde orbital environment.

\begin{figure}
    \centering
    \includegraphics[width=\linewidth]{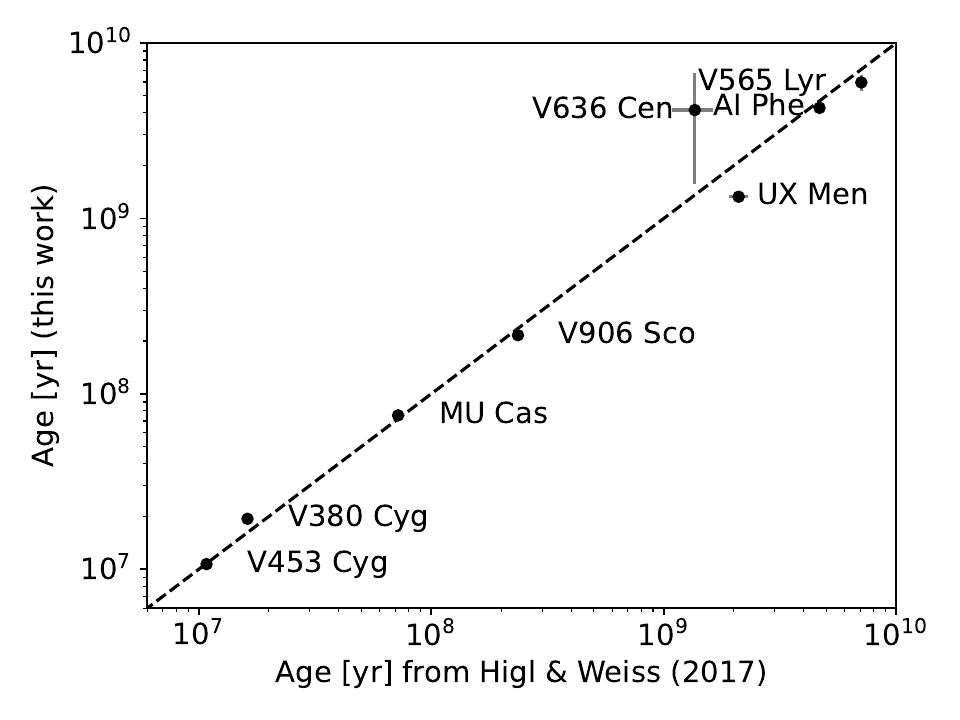}
    \caption{Comparison of our calculated ages with eight systems in common with \citet{higl2017}. Error bars are always present but sometimes smaller than the star symbol size.}
    \label{fig:comp_age}
\end{figure}
    
\begin{figure*}
   \includegraphics[width=0.49\linewidth]{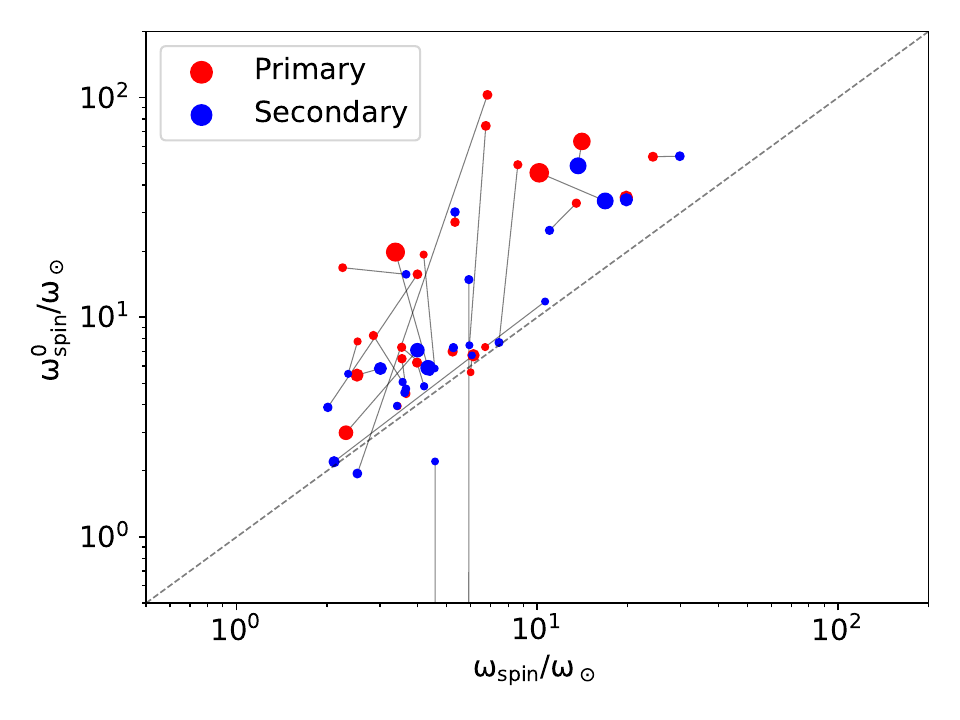}
    \includegraphics[width=0.49\linewidth]{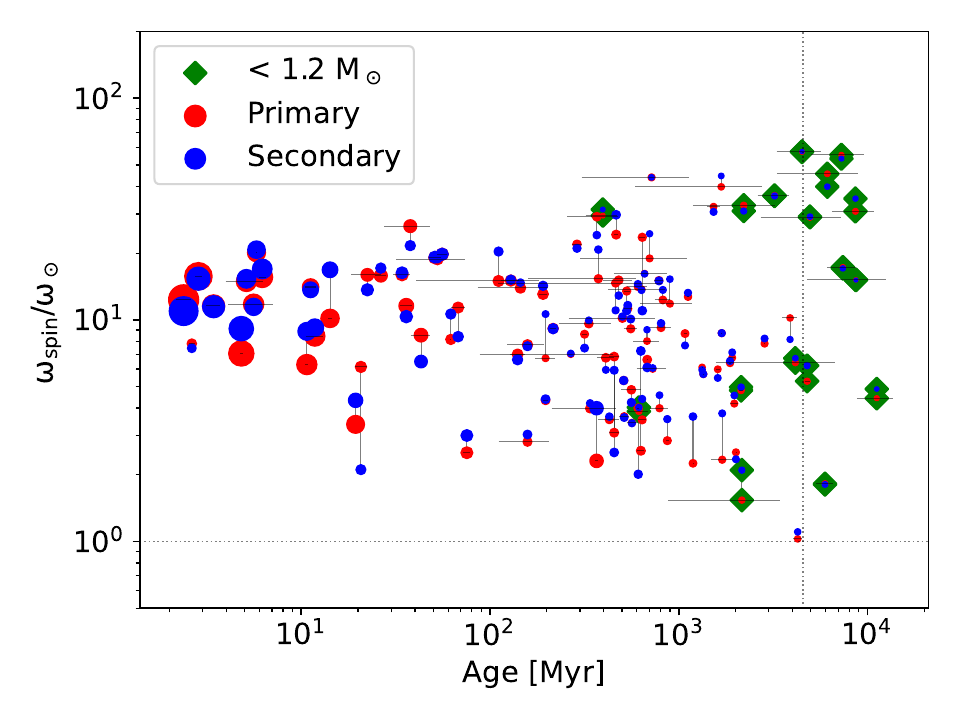}
\caption{Comparison between ZAMS and nowadays angular velocities (left). Nowadays angular velocity as a function of age (right). The size of the dots scales with the mass.}
\label{fig:velocity}
\end{figure*}

\subsection{Discussion}
\label{sec:Discussion}

\subsubsection{Individual System Peculiarities}
\begin{figure*}
    \includegraphics[width=0.49\linewidth]{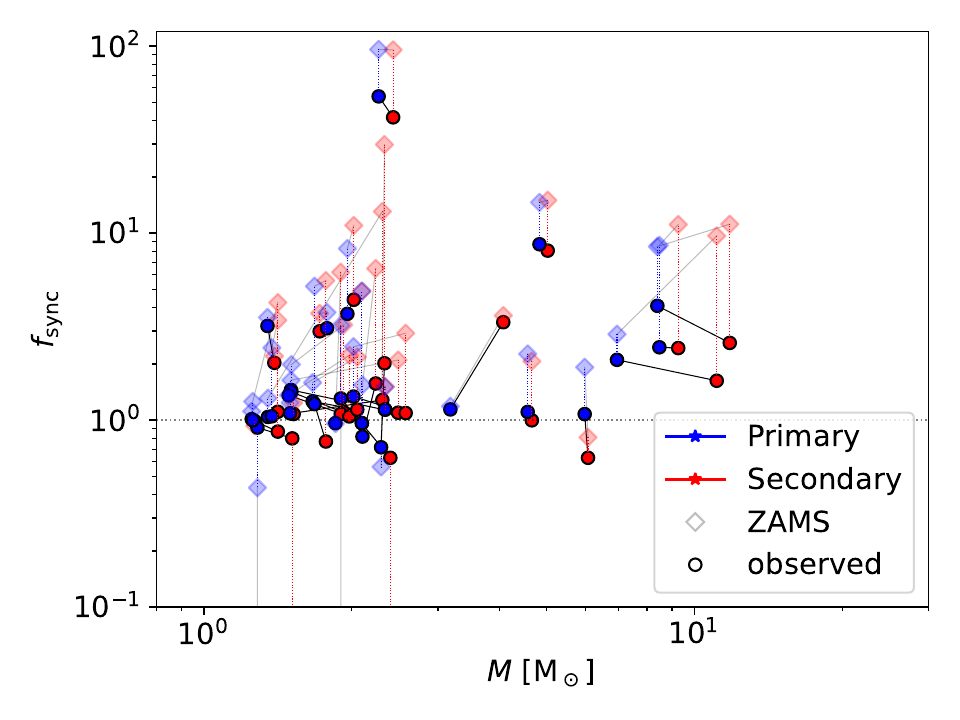}
    \includegraphics[width=0.49\linewidth]{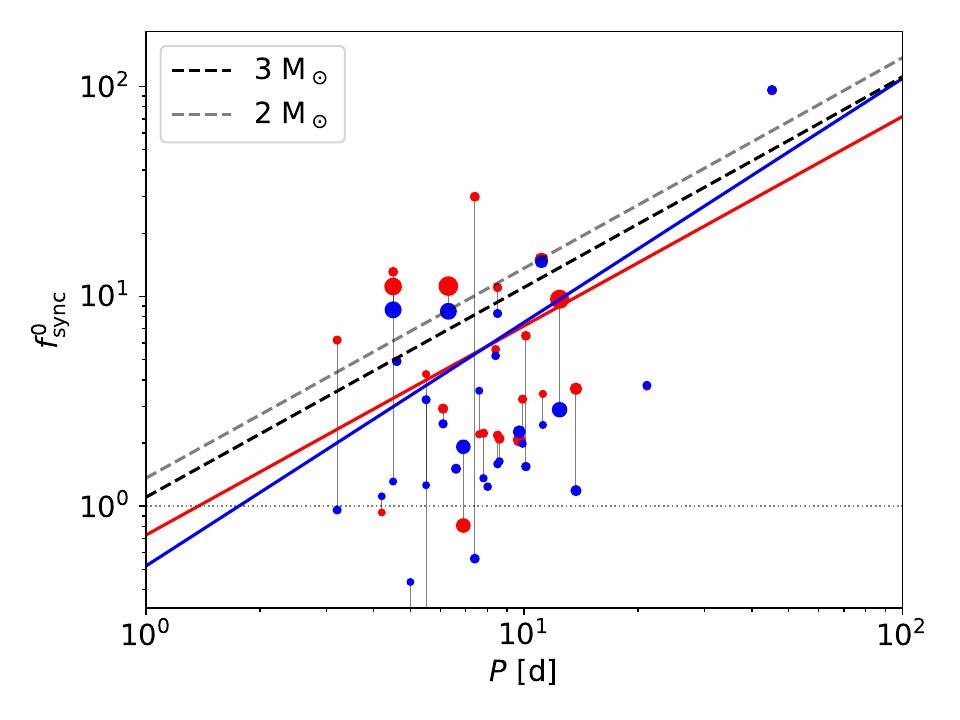}
\caption{ Left: Synchronisation factor at ZAMS (diamonds) and now (circle) as a function of mass for primaries (red) and secondaries (blue). 
Right: synchronisation factor at ZAMS as a function of orbital period. Theoretical values for a binary of (3+2)~M$_\odot$ are shown in black and grey dashed lines, assuming a ZAMS equatorial velocity of 100 km/s.}
\label{fig:fsync}
\end{figure*}

The results obtained in this study, as presented in Tables~\ref{tab:age} and~\ref{tab:sync}, are visualized in Figs.~\ref{fig:velocity} and~\ref{fig:fsync}. 
Our calculations show that the two components of AI Phe have exhausted the core hydrogen, as well as primaries of V432 Aur, SZ~Cen and V380~Cyg. In addition, we found that V1388~Ori is semi-detached because the primary has started to transfer mass onto the secondary. From the mass of the primary (7.42~\Msun), the mass ratio ($q=0.70$) and the orbital period (2.2~d) we estimate a filling factor above 0.9 which can explain the fact that the light curve exhibits ellipsoidal variations in addition to eclipses (see, e.g., the TESS light curve analyzed by \citealt{southworth2022}). However, this system has been classified as detached by \citet{williams2009} and \citet{southworth2022} Consequently, our finding that V1388~Ori is semi-detached should be treated with caution. Our age determined for this star of $36\pm2.7$~Myr is in good agreement with the $33\pm2$~Myr estimate of \citet{southworth2022}.

\subsubsection{Comparison of Age Determinations}

We compare our age determinations with those of \citet{higl2017}, who employed the stellar evolution code \texttt{Garstec} \citep{weiss2008}, in Fig.~\ref{fig:comp_age}. Of the eight systems common to both studies, our age estimates are in good agreement, with the notable exception of UX~Men. For this system, \citet{higl2017} incorporated atomic diffusion (the gravitational settling of heavy elements) and noted that convective overshooting was negligible because UX~Men is in the very early stages of its main-sequence evolution, where the convective core has not yet reached its maximum extent. In contrast, our models account for overshooting from convective regions but neglect atomic diffusion. The discrepancy in the age of UX~Men is likely dominated by the differences in the treatment of overshooting.

Additionally, our age estimate for V636~Cen remains consistent with that of \citet{higl2017} within 2$\sigma$. However, we note that their model for V636~Cen requires a spot coverage of one-third of the stellar surface to match observations.
This value appears unrealistically high for the typical magnetic activity levels of G- and K-type stars \citep[e.g.][]{2005LRSP....2....8B,2009A&ARv..17..251S}, which generally exhibit much lower spot filling factors unless they are in states of extreme RS~CVn-like activity.

\subsubsection{Spin Evolution and Synchronisation Factors}
In the left panel of Fig.~\ref{fig:velocity}, we show the estimated ZAMS angular velocities $\omega_\mathrm{spin}^0$ as a function of the currently observed angular velocities $\omega_\mathrm{spin}$ for primary stars (in red) and secondary stars (in blue). We use as reference a solar angular momentum $\omega_\odot=2.865\times10^{-6}$~rad~s$^{-1}$. The less massive secondary stars are more frequently synchronized or closer to synchronisation compared to the more massive primary stars which store more spin angular momentum. Systems with the largest differences in ZAMS angular velocities between components are V364~Lac and YZ~Cas.

In the right panel of Fig.~\ref{fig:velocity}, we plot the present-day measured angular velocities as a function of the computed ages. The Sun is located at the intersection of the dotted lines. The components of the youngest and more massive systems tend to have angular velocities approximately ten times higher than that of the Sun. The system AI Phe has an age and observed angular velocity close to solar values (4.27 Gyr, with primary $\omega_1 = 1.03$ and secondary $\omega_2 = 1.11$ $\omega_\odot$). AI Phe is the second system (after $\delta$ Vel) with the longest orbital period of 24.6 days in our sample, indicating that the mutual influence between its components is among the weakest.  Without strong tidal forces forcing it to maintain a high rotational speed, AI Phe has been able to naturally slow down over its 4.27 Gyr lifespan, resulting in an angular velocity highly comparable to our Sun.

The rotational evolution of our sample shows a distinct departure from the behavior observed in single-star populations. While models for isolated low-mass stars, such as those by \citet{gallet2015}, predict a significant narrowing of the angular velocity dispersion as stars age beyond 1~Gyr due to magnetic braking, our binary sample maintains a broad range of velocities. Specifically, for our stars with $M \leq 1.2$~\Msun\ older than 1~Gyr, we observe angular velocities ranging from one to 60 times the solar value (see Fig.~\ref{fig:velocity}, right panel). This persistent dispersion, contrasting with the convergence seen in single-star clusters of similar ages, highlights the dominant role of tidal interactions. In these systems, tidal coupling effectively counteracts the deceleration typically induced by magnetic braking, maintaining higher rotation rates throughout the main-sequence evolution.

In the left panel of Fig.~\ref{fig:fsync}, we present the present-day synchronization factors for the primary (red) and secondary (blue) components as a function of stellar mass. Black lines connect the components of each system, while dashed lines trace the evolutionary path from the ZAMS to the current state. We identify two primary trends in the sample's rotational evolution: (i) regardless of the initial state at ZAMS, whether the components are initially over-synchronized, under-synchronized, or, as in the cases of HD~71636 and V396~Cas, rotating in a retrograde sense, there is a universal convergence toward synchronization ($f_{\mathrm{sync}} \approx 1$); (ii) the degree to which a system has approached this state provides a direct probe of the competition between tidal timescales and the nuclear lifetime of the components. While this convergence is the expected outcome of the tidal physics implemented in our models, the value of this result lies in the varying degrees of completion observed across the sample. The fact that several systems remain significantly out of synchronization despite these forces suggests that either the tidal coupling is relatively weak for their specific orbital configurations, or that the systems are sufficiently young that the tidal torque has not yet dominated the initial angular momentum distribution.

The right panel of Fig.~\ref{fig:fsync} displays the synchronization factor at the ZAMS ($f_{\mathrm{sync}}^0$) as a function of the orbital period, with symbol sizes proportional to stellar mass. As expected from tidal theory, the degree of synchronization decreases sharply with increasing orbital period. To quantify this, we performed a linear regression in the $\log{f_{\mathrm{sync}}^0}$--$\log{P}$ plane, yielding slopes of approximately $0.7$ for primaries and $1.1$ for secondaries. These empirical slopes are consistent with the theoretical scaling derived from the definition of the synchronization factor. Since $f_{\mathrm{sync}} = v_{\mathrm{eq}} / v_\mathrm{eq}^\sync$ and the synchronized velocity scales with the period as $v_\mathrm{eq}^\sync = 2\pi R / P$, it follows that for a fixed initial equatorial velocity $v_{\mathrm{eq}}^0$, the synchronization factor scales linearly with the period ($f_{\mathrm{sync}}^0 \propto P$). Variations in the assumed $v_{\mathrm{eq}}^0$ (here taken as $100$~km\,s$^{-1}$) or stellar radii result in a vertical shift of the distribution but do not alter this fundamental power-law slope. The slight deviations from a slope of one in our sample reflect the internal distribution of stellar radii and initial velocities across the different mass regimes.

\begin{figure}
    \centering
    \includegraphics[width=\linewidth]{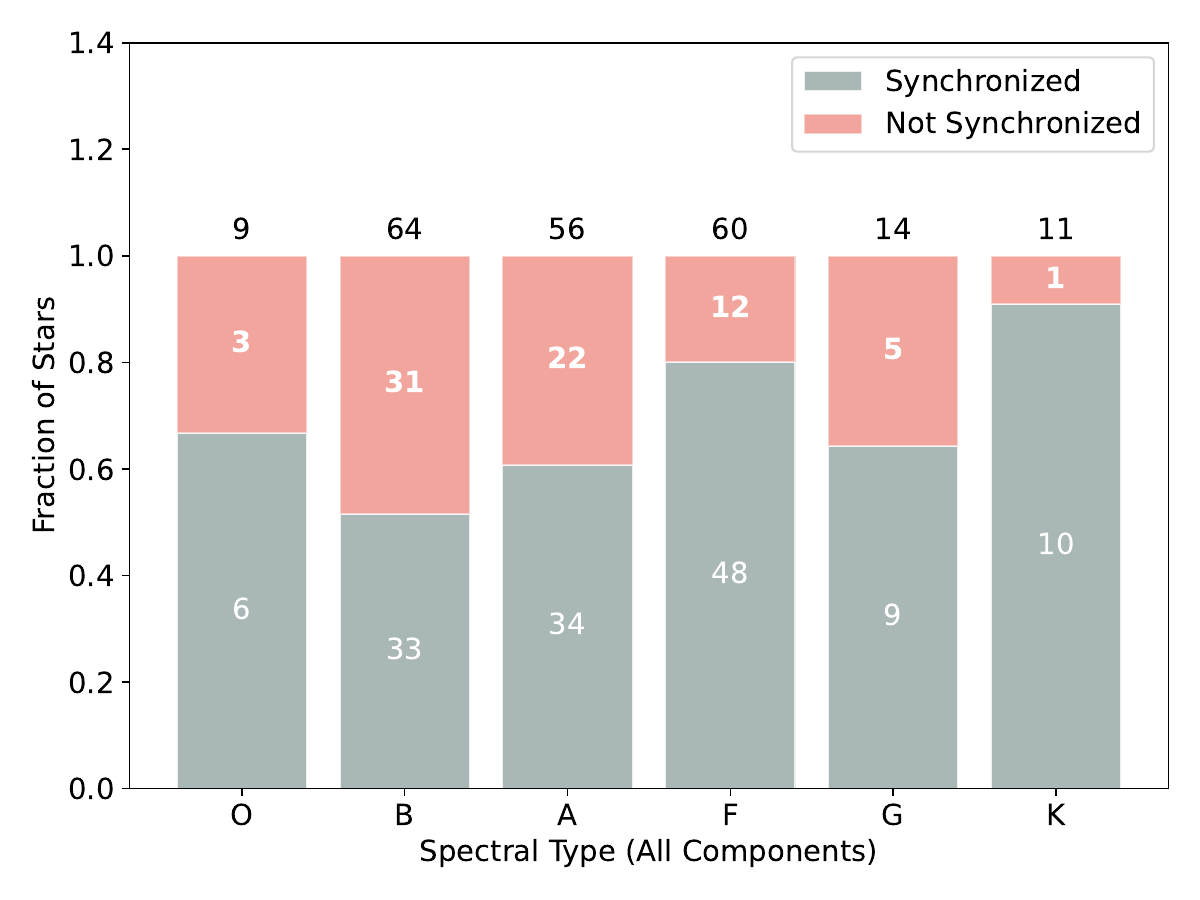}
    \caption{Cumulative star counts where each spectral type bar is normalized to unity. The number of unsynchronised systems (red) decreases with spectral type, \emph{i.e.} with decreasing effective temperature because later spectral type have longer lifetimes.}
    \label{fig:sync_st_histo}
\end{figure}

Following \citet{hobson-ritz2025}, we illustrate the fraction of non-synchronized stars relative to synchronized ones in Fig.~\ref{fig:sync_st_histo}, adopting their synchronization criterion of $0.8 < f_{\mathrm{sync}} < 1.2$. This distribution can be compared to their figure~14. While we find a higher overall proportion of non-synchronized systems, this likely stems from selection effects: our sample is roughly four times smaller and is dominated by BAF-type stars, whereas the \citet{hobson-ritz2025} sample, derived from the TESS EB catalogue \citep{prsa2022}, is dominated by FGK types. Furthermore, our sample includes periods up to 45~d, whereas theirs is restricted to $P < 15$~d. Given that tidal torques scale strongly with the inverse of the orbital period, the systems in their study experience significantly stronger tidal forces, qualitatively explaining their higher synchronization fractions.
We also observe a clear correlation between spectral type and synchronization, with the synchronized fraction increasing toward later types. This trend is driven by two primary physical factors. First, the efficiency of tidal dissipation is significantly higher in later-type stars (FGKM) due to the presence of deep outer convective zones, which provide more effective damping compared to the radiative envelopes of O and B stars. Second, the longer main-sequence lifetimes of late-type stars allow for tidal interactions to act over much greater timescales. 
For B-type stars (our most robust subsample with 64 stars) we find that approximately 50\% are non-synchronized, in good agreement with the findings of \citet{hobson-ritz2025}. In contrast, for late G and K stars, our non-synchronized fraction is approximately 25\%, compared to roughly 10\% in their study. This comparison confirms that rotational synchronization is highly dependent on spectral type; later-type stars achieve synchronism far more frequently than their more massive counterparts, reflecting the combined influence of enhanced tidal dissipation and longer evolutionary timescales.
Among the later-type stars, systems that remain unsynchronized are predominantly associated with wider orbital separations. The unsynchronized subsets of F-, G-, and K-type stars exhibit mean orbital periods more than twice as long as those of their synchronized counterparts (e.g., 9.3 d compared to 4.2 d for F-type systems).  
Longer orbital periods remain the primary factor preventing complete synchronization in the later-type sample.

\subsubsection{Tidal Theories}
Several studies have debated the validity and applicability of theories describing tidal interactions in binary systems. \citet{rieutord1997} argued that the meridional circulation theory proposed by \citet{tassoul1987,tassoul1988} significantly overestimates tidal damping in binaries. In response, \citet{tassoul1997} defended their theory, pointing out that \citet{rieutord1997} did not account for the lack of axial symmetry in non-synchronous rotators, which is the source of large-scale meridional flow. The counter-examples involving Jupiter and Io, or 51 Peg and its Jovian planet Dimidium, cited by \citet{rieutord1997}, are not directly applicable, as the Tassoul formalism was specifically developed for \emph{stellar} binaries, not systems involving planets or satellites with extreme mass ratios. Among others, \citet{claret1995} incorporated the Tassoul theory into their work on binaries.
Criticisms have also been directed at the Zahn formalism for treating tidal effects in binaries. For instance, \citet{sciarini2024} reported inconsistencies in the implementation of Zahn's prescription. \citet{tomkin2006} noted that the two main theories of orbital circularization and rotational synchronisation (\citealt{zahn1977, tassoul1992}) significantly disagree on absolute timescales but concur that synchronisation should occur before circularization in systems like RR Lyn, an EB and SB2 consisting of an A7 and an F0 star.
\citet{lacy2004} found that the synchronisation and circularization timescales for V885 Cyg align with the Tassoul theory. However, \citet{southworth2005} reported conflicting timescales in the detached eclipsing binary WW Aur, challenging both the Tassoul and Zahn tidal theories. To date, no clear consensus has emerged on the most realistic formalism for treating tidal effects on synchronisation and circularization. An insightful review of tidal interactions in close binary stars is provided by \citet{mazeh2008}.
Furthermore, recent developments employing a non-perturbative approach to tidal interactions indicate that the traditional perturbative approach may underestimate tidal interactions between stellar components by up to ~40\% for close low-mass binaries \citep{fellay2024}.

In addition, improving the determination of $E_2$ is an area of active refinement. For stars with radiative envelopes and convective cores, the synchronisation timescale in Zahn's formalism depends sensitively  on the tidal torque constant, $E_2$ (see Eq.~\ref{eq:tsync_RE}). In this work, we adopted the tabulated values of \citep{claret2004}. However, more recent studies \citep{2010ApJ...725..940Y,2018A&A...616A..28Q} have shown that $E_2$ is highly sensitive to the internal stellar structure, providing updated prescriptions that depend explicitly on the fractional radius of the convective core. Implementing such dynamically evolving $E_2$ formulations lies beyond the scope of the present binary evolution models; nevertheless, their inclusion in future work may yield more accurate synchronization timescales for the massive components of our sample.

\section{Conclusion} 
In this study, we determined the ages and initial rotational velocities of stars in detached eclipsing binaries from the catalogue of \citet{eker2014}. We identified 108 binaries suitable for this analysis by selecting those with well-defined fundamental parameters (i.e., having defined $M$, $R$, $L$, $\Teff$, and \veq\ for both components) and excluding systems with an M-type component. Among our sample, we found that V1388~Ori is currently semi-detached, yet its age could still be determined reliably.
The derived ages for all 108 binaries, along with their present hydrogen core mass fractions, are provided in Table~\ref{tab:age}. Notably, both components of AI~Phe, as well as the primaries of SZ~Cen, V432~Aur, and V380~Cyg, have exhausted their core hydrogen. This highlights our methodology's ability to track binary evolution and constrain ages even as stars reach the terminal-age main sequence and begin significant structural expansion.

For 29 of these systems, we successfully determined the initial equatorial velocities at ZAMS, as detailed in Table~\ref{tab:sync}. These results reveal an increasing dispersion of spin velocities with age, and demonstrate that the computed initial synchronisation factors of primaries and secondaries reproduce well their expected increases with orbital period.
The remaining 79 binaries currently rotate (almost) synchronously. Under our working assumption of circular orbits, this implies their ZAMS velocities cannot be uniquely constrained, as a wide range of initial values would naturally evolve to the observed synchronous state. However, we caution that 27\% of our total sample possesses an orbital eccentricity $e>0.1$. Because eccentricity was neglected in this study, future models incorporating eccentric orbital evolution may yield different initial spin constraints for these specific systems.

For the asynchronous systems in our sample, we identified a unique ZAMS spin velocity required to reproduce present-day observations. We identify three distinct evolutionary pathways: the majority of systems originate in an over-synchronized state and subsequently spin down; a smaller subset is born under-synchronized and spins up; and a rare group, including HD~71636 and V396~Cas, likely originated with retrograde spins. While tidal torques consistently drive the synchronization factor toward unity, stellar evolution provides a competing effect. As a star approaches the end of the main sequence, the increase in its radius leads to a higher required synchronous velocity ($v_{\mathrm{sync}} \propto R$). This expansion effectively acts to drive $f_{\mathrm{sync}}$ below unity, even if the absolute rotational velocity remains relatively stable.

The persistence of high rotation rates in our older binary systems suggests that tidal interactions effectively counteract the standard rotational braking observed in single stars. Although our current binary models do not explicitly include magnetic braking through magnetized winds, the fact that our sample, particularly the later-type stars, remains far more synchronized than equivalent single-star populations demonstrates that tidal coupling provides a significant reservoir of angular momentum. This coupling replenishes the angular momentum lost via winds, thereby preventing the dramatic spin-down typically governed by the Skumanich-like laws in isolated stars.

\begin{acknowledgements}
We thank the referee for the useful comments and suggestions that improve the quality of the manuscript.
T.M. is granted by the BELSPO Belgian federal research program FED-tWIN under the research profile Prf-2020-033\_BISTRO. LS is FNRS research director.
\end{acknowledgements}

\bibliographystyle{aa}
\bibliography{aa60757-26}

\end{document}